\documentclass[11pt]{article}

\usepackage[letterpaper,margin=1in]{geometry}
\usepackage[T1]{fontenc}
\usepackage[utf8]{inputenc}
\usepackage{lmodern}
\usepackage{microtype}

\usepackage{cite}
\usepackage{url}
\usepackage{amsmath,amssymb,amsfonts}
\usepackage{graphicx}
\usepackage{textcomp}
\usepackage{threeparttable}
\usepackage{booktabs}
\usepackage{algorithmic}
\usepackage{multirow}
\usepackage{siunitx}
\usepackage{algorithm}
\usepackage[dvipsnames]{xcolor}
\usepackage[caption=false,font=footnotesize]{subfig}
\usepackage{tabularx}
\usepackage{adjustbox}
\usepackage{float}
\usepackage{authblk}
\usepackage[colorlinks=true,linkcolor=blue,citecolor=blue,urlcolor=blue]{hyperref}

\usepackage{pifont}

\usepackage{bm}

\def\BibTeX{{\rm B\kern-.05em{\sc i\kern-.025em b}\kern-.08em
    T\kern-.1667em\lower.7ex\hbox{E}\kern-.125emX}}

\title{\mbox{RAPID-DNN}: Reliability-Aware Partitioning for Robust DNN Inference Deployment on Edge AI}

\author[1]{Mukta Debnath}
\author[2]{Krishnendu Guha}
\author[1]{Debasri Saha}
\author[1,3]{Amlan Chakrabarti}
\author[4]{Susmita Sur-Kolay}

\affil[1]{A.K. Choudhury School of Information Technology, University of Calcutta, Kolkata 700073, India}
\affil[2]{School of Computer Science and Information Technology, University College Cork, Ireland}
\affil[3]{Indian Institute of Technology, Kharagpur, India}
\affil[4]{Indian Statistical Institute, Kolkata, India}

\date{}

\begin{document}

\maketitle

\begin{abstract}
% Deep neural networks (DNNs) deployed on heterogeneous edge AI
% accelerators are increasingly susceptible to runtime hardware faults
% caused by process variation, voltage fluctuations, timing violations,
% device aging, and environmental disturbances. 
Deep neural networks (DNNs) deployed on heterogeneous edge accelerators are increasingly susceptible to runtime hardware faults. 
In heterogeneous edge deployments, stringent power and resource constraints often necessitate aggressive voltage scaling and limit the use of hardware fault mitigation techniques.
Process variation, thermal stress, and device aging further increase the susceptibility of edge accelerators to runtime faults, making reliable DNN inference a key deployment challenge.
These faults manifest as
transient bit-level corruptions that propagate through DNN inference,
resulting in significant degradation of prediction accuracy. 
Existing
DNN partitioning frameworks primarily optimize latency and energy under
fault-free assumptions, producing deployment strategies that often fail
to maintain reliable inference under realistic operating conditions.
This paper presents RAPID-DNN, a reliability-aware DNN partitioning
framework for heterogeneous edge accelerators that explicitly incorporates
runtime reliability into the deployment optimization process.
Unlike existing partitioning frameworks that optimize only latency and energy, RAPID-DNN introduces runtime reliability as a third optimization objective through layer-wise fault sensitivity modeling and multi-objective deployment optimization. It first
performs layer-wise reliability characterization through systematic
fault injection in the weight and activation domains to identify
vulnerability-critical layers. The resulting sensitivity profile guides
a three-objective NSGA-II optimization framework that jointly minimizes
inference latency, energy consumption, and expected fault-induced
accuracy degradation.
% RAPID-DNN first performs layer-wise reliability characterization through
% systematic fault injection in the weight and activation domains to
% quantify the fault-induced inference accuracy degradation associated
% with each DNN layer. 
% The resulting sensitivity profile identifies vulnerability-critical layers and guides reliability-aware partition optimization. 
% Candidate partitions are evaluated using hardware performance profiles within a three-objective NSGA-II framework that jointly minimizes inference latency, energy consumption, and expected fault-induced accuracy degradation.
Experimental evaluation on AlexNet, SqueezeNet, ResNet18, VGG16, and
MobileNetV2 using heterogeneous Eyeriss and SIMBA accelerator profiles
demonstrates that RAPID-DNN consistently improves inference robustness
over the fault-agnostic baseline across diverse fault models and
injection probabilities.
Across the evaluated benchmarks under the representative mixed random
fault configuration, RAPID-DNN achieves an average improvement of 9.81 percentage points in Top-1 inference accuracy while reducing the average
expected fault-induced accuracy degradation by 36.86\%, incurring only
2.38\% average latency overhead and 3.34\% average energy overhead
relative to the fault-agnostic baseline.
These results demonstrate that incorporating reliability into DNN partition optimization enables robust deployment of AI applications on heterogeneous edge accelerators under runtime hardware faults.

\end{abstract}


% ============================================================
% ============================================================
\section{Introduction}
\label{sec:intro}
% ============================================================

Deep Neural Networks (DNNs) have become a fundamental component of
modern edge intelligence systems, enabling real-time inference in
applications such as autonomous driving, industrial automation, medical
diagnostics, smart surveillance and intelligent IoT
platforms~\cite{cnnparted,edgent,neurosurgeon}. These deployments must satisfy strict
constraints on latency, energy consumption, and runtime reliability while
operating under limited computational and power budgets. Thus, DNN inferencing is increasingly partitioned across
heterogeneous edge accelerators including CPUs, GPUs, FPGAs and
domain-specific neural processing units (NPUs) within System-on-Chip
(SoC) architectures~\cite{SPINN,roadrunner,MAESTRO}.

Existing DNN partitioning frameworks such as CNNParted~\cite{cnnparted}, Edgent~\cite{edgent}, Neurosurgeon~\cite{neurosurgeon} and RoaDRuNNer~\cite{roadrunner} primarily optimize layer-to-device mappings for latency, throughput and energy efficiency. However, these approaches generally assume fault-free hardware execution and therefore overlook the impact of runtime hardware faults on inference reliability. Recent surveys on DNN reliability assessment and AI fault injection have identified hardware reliability as a critical challenge for edge AI deployment, emphasizing the growing need for reliability-aware design methodologies~\cite{Yu2025Survey,Ahmadilivani2024SLR, ref22, ref23}. Modern edge accelerators operate under aggressive voltage scaling, high integration density, process variation, thermal stress and device aging, making them increasingly susceptible to transient runtime faults.

These hardware faults typically manifest as bit-level corruptions in
weights, activations, feature maps, and intermediate tensor computations
during inference execution. Prior studies have shown that such faults may
originate from naturally occurring phenomena including voltage
fluctuations, timing violations, single-event upsets (SEUs), memory
instability, and electromagnetic disturbances~\cite{Ahmadilivani2024SLR, fi-cnn1,sifi, ref27}. 
Although this work focuses on naturally occurring transient faults, similar bit-level corruptions may also arise under fault-injection attacks such as voltage glitching, electromagnetic
interference and Rowhammer-style memory disturbance
attacks~\cite{FIDNN,heterosplit}. Regardless of origin, these runtime faults
can silently propagate through DNN computations and significantly degrade
prediction accuracy without modifying the model architecture or
retraining the network.

The reliability challenge becomes considerably more complex in
heterogeneous edge deployments because different accelerators differ in architecture, memory hierarchy, dataflow and execution characteristics, leading to different fault propagation behavior. A fault introduced in one layer may
propagate and amplify through subsequent tensor computations, leading to
large-scale inference degradation. This
behavior is further influenced in partitioned DNN execution by inter-device communication and
partition boundaries. 
A partitioning strategy may either
attenuate or amplify the propagation of hardware faults, depending on the placement of partition boundaries and the
vulnerability of individual layers.
Consequently, a partition strategy
that appears optimal from a latency-energy perspective may become highly
unreliable if vulnerability-critical layers are mapped to vulnerable hardware
resources.

Despite extensive research on DNN partitioning and hardware reliability
independently, their interaction in heterogeneous edge deployments
remains insufficiently explored. This paper addresses this gap through
RAPID-DNN (\textbf{R}eliability-\textbf{A}ware
\textbf{P}artitioning for Robust DNN \textbf{I}nference
\textbf{D}eployment on Edge AI), a framework for deploying DNNs on
heterogeneous edge accelerators under runtime hardware faults.
RAPID-DNN combines layer-wise reliability characterization with
sensitivity-guided multi-objective partition optimization. Candidate
partitions are evaluated using hardware performance profiles within a
three-objective NSGA-II~\cite{nsga2} framework that jointly minimizes
inference latency, energy consumption, and expected fault-induced
accuracy degradation.

The principal contributions of this work are summarized as follows.

\begin{itemize}

\item \textbf{Reliability-aware DNN partitioning framework:}
We propose RAPID-DNN, a framework for reliability-aware partitioning of
DNN inference across heterogeneous edge accelerators. Unlike
conventional partitioning approaches that optimize only latency and
energy, RAPID-DNN explicitly considers expected fault-induced accuracy
degradation during deployment planning, enabling robust partition
selection under runtime hardware faults.

\item \textbf{Layer-wise reliability characterization:}
We introduce a stochastic fault-injection methodology in the weight and
activation domains for layer-wise reliability characterization. The
resulting sensitivity profile identifies vulnerability-critical layers
used during partition optimization.

\item \textbf{Layer-wise sensitivity-guided candidate partition evaluation:}
We propose a sensitivity-guided evaluation strategy that combines
hardware performance profiles with the identified layer sensitivities to
estimate the expected fault-induced accuracy degradation of candidate
deployment configurations.
By prioritizing the most sensitive layers during stochastic fault
injection, the proposed evaluation efficiently estimates the reliability
of candidate partitions while reducing computational cost.

\item \textbf{Comprehensive experimental evaluation:}
We comprehensively evaluate RAPID-DNN using five widely adopted CNN benchmarks—AlexNet, SqueezeNet, ResNet18, VGG16, and MobileNetV2, which represent a diverse range of network architectures, model complexities, and deployment characteristics commonly used in edge AI research. The evaluation is conducted on heterogeneous Eyeriss and SIMBA accelerator profiles under various runtime fault scenarios.
Experimental results demonstrate an average improvement of 9.81 percentage points in Top-1 inference accuracy and  36.86\%
reduction on an average in expected fault-induced accuracy degradation, while
incurring only 2.38\% average latency overhead and 3.34\% average
energy overhead relative to the fault-agnostic CNNParted baseline.
The evaluation includes 
robustness analysis under varying fault injection probabilities and
bit-level fault models, latency-energy-reliability trade-off
analysis, optimization convergence, and ablation studies.

\end{itemize}

The remainder of this paper is organized as follows.
Section~\ref{sec:background} reviews  related works and identifies
the underlying research gap.
Section~\ref{sec:threat_model} presents the threat model and fault
characterization.
Section~\ref{sec:problem} formulates the optimization problem.
Section~\ref{sec:method} describes the proposed methodology.
Section~\ref{sec:experiments} presents the experimental setup, evaluation results, ablation
analysis, and a discussion of the limitations of the proposed framework.
Finally, Section~\ref{sec:conclusion} concludes the paper and outlines future research directions.

\section{Related Works}
\label{sec:background}

\begin{table*}[htbp]
\caption{Comparison of representative DNN partitioning and reliability
frameworks with the proposed RAPID-DNN framework.}
\label{tab:relwork}
\centering
\renewcommand{\arraystretch}{1.25}
\setlength{\tabcolsep}{3pt}
\begin{adjustbox}{max width=\textwidth}
\begin{tabular}{|p{2.2cm}|p{2.8cm}|p{1.2cm}|p{1.6cm}|p{1.6cm}|p{1.6cm}|p{2.4cm}|p{1.5cm}|}
\hline
\textbf{Framework} &
\textbf{Optimization Strategy} &
\textbf{Fault-Aware} &
\textbf{Runtime Fault Injection} &
\textbf{Layer-Wise Reliability Characterization} &
\textbf{Sensitivity-Guided Partition Evaluation} &
\textbf{Target Platform} &
\textbf{Optimization Objectives} \\
\hline

CNNParted~\cite{cnnparted} &
NSGA-II-based Pareto partitioning &
No & No & No &
No &
Edge accelerators &
Latency, energy \\
\hline

Edgent~\cite{edgent} &
RL-based adaptive DNN partitioning &
No & No & No &
Yes\textsuperscript{\emph{a}} &
Edge--cloud systems &
Latency \\
\hline

Neurosurgeon~\cite{neurosurgeon} &
Profiling-based layer placement &
No & No & No &
No &
Mobile--cloud platforms &
Latency, energy \\
\hline

SPINN~\cite{SPINN} &
Hardware-aware scheduling and code generation &
No & No & Partial &
No &
Custom accelerators &
Latency, energy \\
\hline

RoaDRuNNer~\cite{roadrunner} &
Automatic partitioning and runtime scheduling &
No & No & No &
No &
FPGA + GPU systems &
Latency, energy \\
\hline

MAESTRO~\cite{MAESTRO} &
Analytical cost and dataflow modeling &
No & No & Partial &
No &
Analytical accelerator models &
Energy, performance \\
\hline

SiFi-AI~\cite{sifi} &
Soft-error fault injection framework &
Yes & No & Partial &
No &
DNN accelerators &
Reliability evaluation \\
\hline

BFA~\cite{bfa} &
Targeted bit-flip attack analysis &
Yes & Yes & Partial &
No &
CPU/ GPU &
Attack effectiveness \\
\hline

\textbf{Proposed RAPID-DNN} &
\textbf{Sensitivity-guided reliability-aware partitioning} &
\textbf{Yes} &
\textbf{Yes} &
\textbf{Yes} &
\textbf{Yes} &
\textbf{Heterogeneous edge accelerators} &
\textbf{Latency, energy, reliability} \\
\hline
\end{tabular}
\end{adjustbox}

\smallskip
\begin{minipage}{\linewidth}\footnotesize
Edgent supports runtime adaptation for latency management only; hardware fault conditions are not considered.\\
\end{minipage}
\end{table*}

The deployment of deep neural networks (DNNs) on heterogeneous edge
platforms has stimulated extensive research on workload partitioning,
accelerator-aware optimization and hardware reliability. In parallel,
recent studies have demonstrated that DNN accelerators are increasingly
susceptible to both naturally occurring hardware faults and
adversarially induced bit-level perturbations. However, these research
directions have largely evolved independently, with limited attention
given to how runtime hardware faults influence DNN partitioning
decisions and deployment strategies in heterogeneous edge environments.

\subsection{DNN Partitioning and Edge Deployment}

Existing DNN partitioning frameworks primarily focus on optimizing
latency, throughput, and energy efficiency across heterogeneous
computing platforms. 
CNNParted~\cite{cnnparted} formulates heterogeneous DNN partitioning as a multi-objective NSGA-II optimization problem that jointly minimizes inference latency and energy consumption using hardware performance estimates obtained from Timeloop~\cite{timeloop} and Accelergy~\cite{accelergy}, two widely adopted analytical tools
for hardware-aware evaluation of DNN accelerators. Although effective for performance-oriented deployment, the framework assumes fault-free execution and therefore does not consider runtime hardware reliability.

Timeloop estimates
execution latency, memory traffic and
resource utilization for different accelerator architectures and
dataflows, whereas Accelergy provides architecture-level energy
estimation based on hardware component models. Together, they enable hardware-aware performance estimation and accelerator design-space exploration, making them a standard profiling infrastructure for performance-driven DNN deployment and partitioning studies.
Similarly, RoaDRuNNer~\cite{roadrunner}
investigates runtime scheduling and deadline-aware task offloading,
whereas Edgent~\cite{edgent} and Neurosurgeon~\cite{neurosurgeon}
propose adaptive edge-cloud partitioning through runtime profiling and
layer placement strategies.

Complementary to these deployment frameworks, analytical tools have been
developed to improve accelerator-aware optimization. MAESTRO~\cite{MAESTRO}
provides a data-centric analytical model for evaluating DNN dataflows
and hardware costs, enabling informed accelerator design and mapping
decisions. SPINN~\cite{SPINN} introduces hardware-aware scheduling and
code generation for progressive DNN inference across device and cloud
resources, while HeteroSplit~\cite{heterosplit} jointly optimizes DNN
partitioning under heterogeneous computation and communication
constraints. Collectively, these studies have significantly advanced
performance-driven DNN deployment. Nevertheless, they generally assume
fault-free hardware execution and do not explicitly incorporate runtime
hardware reliability into the partitioning process.

% \subsection{Fault Modeling and Reliability Analysis in DNNs}
\subsection{Fault Injection and Reliability Analysis in DNN}

The reliability of DNN inference under bit-level hardware faults has received
considerable attention with the growing adoption of AI workloads on
resource-constrained edge accelerators. Fault injection techniques are
widely employed to evaluate the effects of transient and permanent
hardware faults by introducing controlled perturbations into weights,
activations, and intermediate computations~\cite{fi-cnn1,fi-cnn2,Ahmadilivani2024SLR,Yu2025Survey}.
SiFi-AI~\cite{sifi} extends this capability through RTL-level fault
simulation supporting bit-level injection across memories, processing
elements, and compute logic, revealing architecture-dependent
vulnerability patterns. Similarly,~\cite{ref24} demonstrates that
inference degradation strongly depends on fault location, bit
significance, and numerical representation.

Existing studies consistently demonstrate that hardware faults affect DNN inference in a highly non-uniform manner, with different layers and numerical representations exhibiting markedly different vulnerability characteristics. In particular, sign-bit and
higher-order bit corruptions typically produce substantially greater
accuracy degradation than least significant bit (LSB)
errors~\cite{bfa,practicalfault}. Furthermore, architectural
characteristics, including memory hierarchy, compute organization, and
dataflow, significantly influence fault propagation and reliability~\cite{chen2019eyeriss,symba}. Despite these insights, most
existing approaches employ fault injection primarily as a post-deployment
evaluation methodology, where robustness is assessed after deployment
rather than incorporated into deployment-time optimization.

\subsection{Adversarial Fault Injection and Hardware Security}

Another line of research investigates adversarial hardware fault injection
through voltage glitching, electromagnetic interference, memory
disturbance attacks and targeted bit-flip attacks
\cite{deepstrike,bfa, ref20, practicalfault}. These studies demonstrate that
carefully induced bit-level perturbations can significantly degrade DNN
inference accuracy or alter model predictions. However, their primary
objective is attack generation, vulnerability analysis and security
evaluation, rather than improving robustness of deployment  under runtime
hardware faults.

Fault attacks have been exploited to reverse engineer neural
network architectures using clock glitches, voltage glitches, laser
fault injection and Rowhammer-style memory disturbances
\cite{Breier2022SNIFF}, further highlighting the security implications
of hardware fault mechanisms.
Complementary to these attack-oriented studies, TensorFI
\cite{Narayanan2023TensorFI} provides a software-level fault injection
framework for TensorFlow applications that enables resilience evaluation
through controlled perturbation of neural network computations,
the objective being reliability assessment instead of
deployment-time optimization.

Recent surveys categorize existing reliability assessment methodologies
into RTL-, architectural-, software-, and application-level fault
injection techniques while emphasizing that the majority of existing
approaches evaluate reliability after deployment rather than
incorporating reliability information into deployment planning and DNN
partition optimization
\cite{Ahmadilivani2024SLR,Yu2025Survey}.

\begin{figure*}[!t]
    \centering
    \includegraphics[
        width=\textwidth,
        height=\textheight,
        keepaspectratio]
    {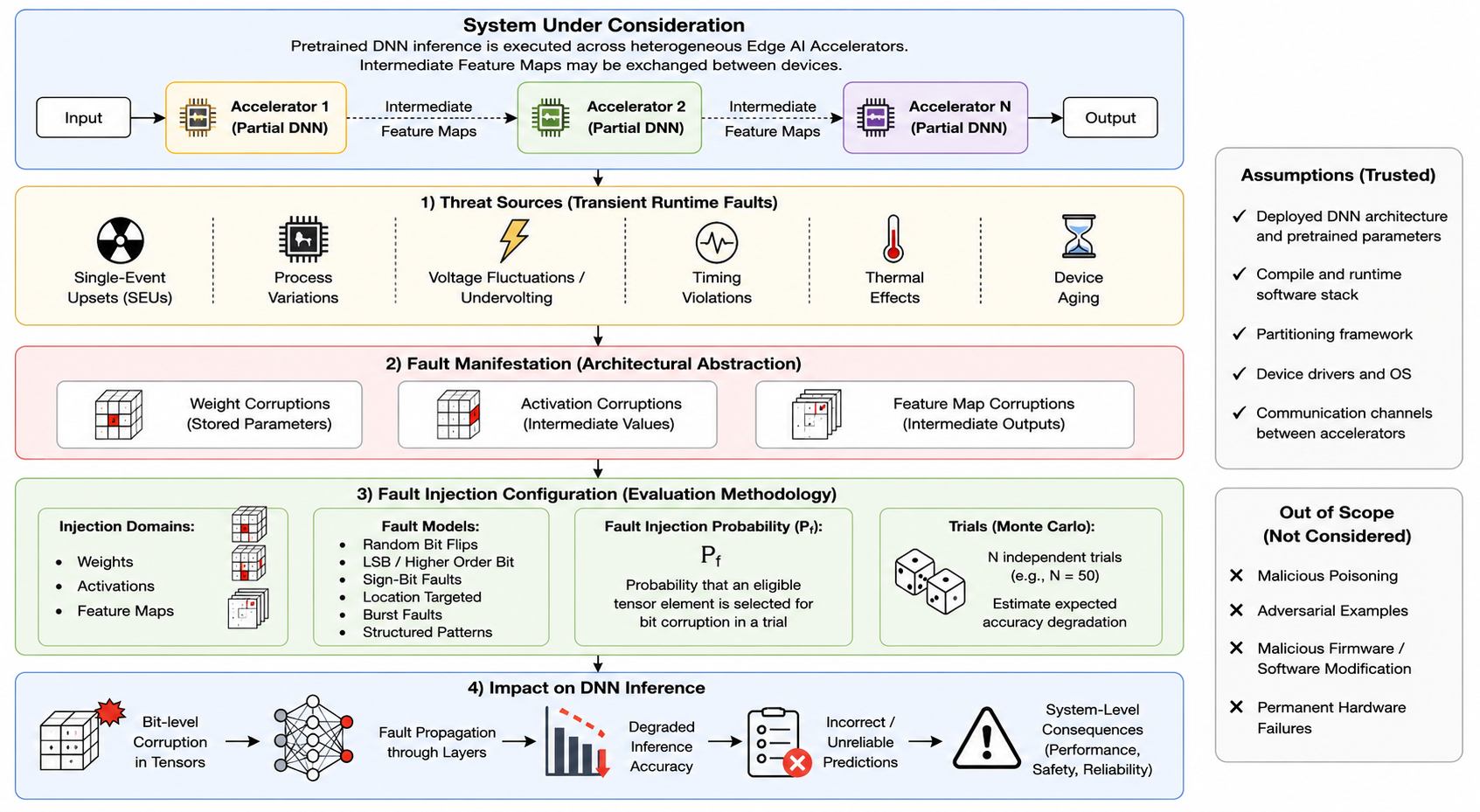}
   \caption{System model, threat model, and fault injection methodology adopted in RAPID-DNN for reliability evaluation of partitioned DNN inference on heterogeneous edge accelerators.}
    \label{fig:fault-model}
\end{figure*}

\subsection{Research Gap}
\label{sec:related_gap}

The comparison in Table~\ref{tab:relwork} highlights the limitations of
existing DNN partitioning and reliability frameworks. Unlike prior
approaches, RAPID-DNN integrates layer-wise reliability
characterization, sensitivity-guided partition evaluation, and
reliability-aware optimization within a unified deployment framework,
enabling joint optimization of latency, energy, and inference
robustness.
The analysis of existing approaches reveals three key research gaps that
remain to be addressed.

\textit{First}, existing DNN partitioning
frameworks~\cite{cnnparted,edgent,neurosurgeon,SPINN,roadrunner,MAESTRO}
primarily optimize deployment with respect to latency and energy while
assuming fault-free hardware execution. Consequently, runtime
reliability is not considered during partition optimization.

\textit{Second}, existing fault injection and reliability
studies~\cite{sifi,fi-cnn1,fi-cnn2,ref24} provide valuable insights into
fault propagation and accelerator vulnerability, but they largely treat
reliability as a post-deployment evaluation problem. Furthermore, they
do not exploit layer-wise vulnerability information to guide deployment
or partitioning decisions.

\textit{Third}, adversarial fault injection
studies~\cite{deepstrike,bfa,practicalfault} primarily focus on attack
generation, security evaluation and vulnerability assessment rather
than developing deployment strategies that improve inference reliability
under transient runtime hardware faults.

These research gaps motivate the reliability-aware DNN partitioning
framework proposed in this work.

\section{System and Threat Model}
\label{sec:threat_model}

\subsection{System Model}

We consider the deployment of pretrained DNNs across heterogeneous edge accelerators that collaboratively execute partitioned DNN inference. The network is partitioned into sequential execution stages mapped onto multiple accelerators, with intermediate feature maps exchanged between devices during inference. The deployment objective is to identify reliability-aware partitioning strategies that preserve inference accuracy while satisfying latency and energy constraints under heterogeneous hardware characteristics.
Figure~\ref{fig:fault-model} illustrates the system model together with the considered threat model and fault injection methodology.

\subsection{Threat Model}

The primary threat considered in this work comprises transient runtime
hardware faults affecting heterogeneous edge accelerators during DNN
inference. Such faults may arise naturally from environmental radiation
(e.g., single-event upsets (SEUs)), process variations, voltage
fluctuations, timing violations caused by aggressive voltage/frequency
scaling, thermal stress during sustained operation, and progressive
device aging. These conditions are commonly encountered in
resource-constrained edge platforms operating under diverse workload and
environmental conditions, where they may induce transient bit-level
corruptions in memories, processing elements, and on-chip data paths.
The resulting bit corruptions can affect DNN weights, activations, or
intermediate feature maps, propagate through subsequent layers, and
ultimately degrade inference accuracy or produce incorrect
predictions.

Hardware fault injection techniques, including voltage glitching,
electromagnetic (EM) fault injection, laser-based fault injection, and
memory disturbance mechanisms (e.g., Rowhammer), are used to emulate
transient bit-level corruptions during
inference~\cite{9316989,FIDNN,deepstrike,bfa,practicalfault,Breier2022SNIFF, 9045134}.
These techniques are employed solely for experimental evaluation and do
not imply the presence of an active adversary.
Runtime faults are abstracted using an architecture-independent fault
model in which bit-level corruptions affect DNN weights, activations,
and intermediate feature maps. This abstraction is widely adopted in
prior DNN reliability
studies~\cite{GUTIERREZZABALLA2024103242, fi-cnn1,sifi,FIDNN,ref24} and enables consistent
evaluation across heterogeneous accelerator architectures.

As illustrated in Figure~\ref{fig:fault-model}, the deployed DNN architecture, pretrained model parameters, compiler,
runtime software stack, and communication framework are assumed to be
trusted. Active adversaries capable of manipulating model parameters,
injecting malicious firmware, compromising accelerator hardware,
crafting adversarial inputs, or inducing deliberate hardware faults are
outside the scope of this work. Likewise, permanent hardware failures
requiring hardware redundancy or replacement are not considered.

%-----------------------------------------------------------

\subsection{Fault Injection Configuration}
\label{sec:fault_config}

Hardware faults are emulated through stochastic bit-level fault
injection during DNN inference. Let $p_f$ denote the fault injection
probability, defined as the probability that an eligible tensor element
is independently selected for bit-level corruption during a fault-
injection trial. For a given fault injection probability $p_f$, each
eligible weight or activation element is independently sampled using a
Bernoulli trial with probability $p_f$. For every selected element,
exactly one bit is corrupted according to the specified fault model
(LSB, higher-order, sign-bit, or random).

Fault injection is independently performed in the weight and activation
domains to capture the different impacts of runtime hardware
faults. 
Weight-domain faults model transient corruption of stored network
parameters during inference, whereas activation-domain faults capture
errors affecting intermediate feature maps and computations.
Following established practice in DNN fault-injection
studies~\cite{HTDNNSurvey,fi-cnn1,sifi,FIDNN}, RAPID-DNN employs
stochastic bit-flip fault injection with configurable fault rates,
including least significant bit (LSB), higher-order bit, and sign-bit
corruptions.

The numerical impact of a runtime fault depends strongly on the
significance of the corrupted bit. In fixed-point DNN
representations, higher-order and sign-bit corruptions generally produce
substantially larger numerical perturbations than LSB faults, resulting
in more severe degradation of inference
accuracy~\cite{bfa,practicalfault,sifi}. 
Each fault configuration is evaluated over repeated independent
fault-injection trials, with fault locations randomly sampled according
to the specified fault injection probability. The expected
fault-induced accuracy degradation is computed as the average across all
trials and serves as the reliability objective in the optimization
problem described in Section~\ref{sec:problem}.

\begin{figure*}[t!]
    \centering
    \includegraphics[width=\linewidth, height=98mm]{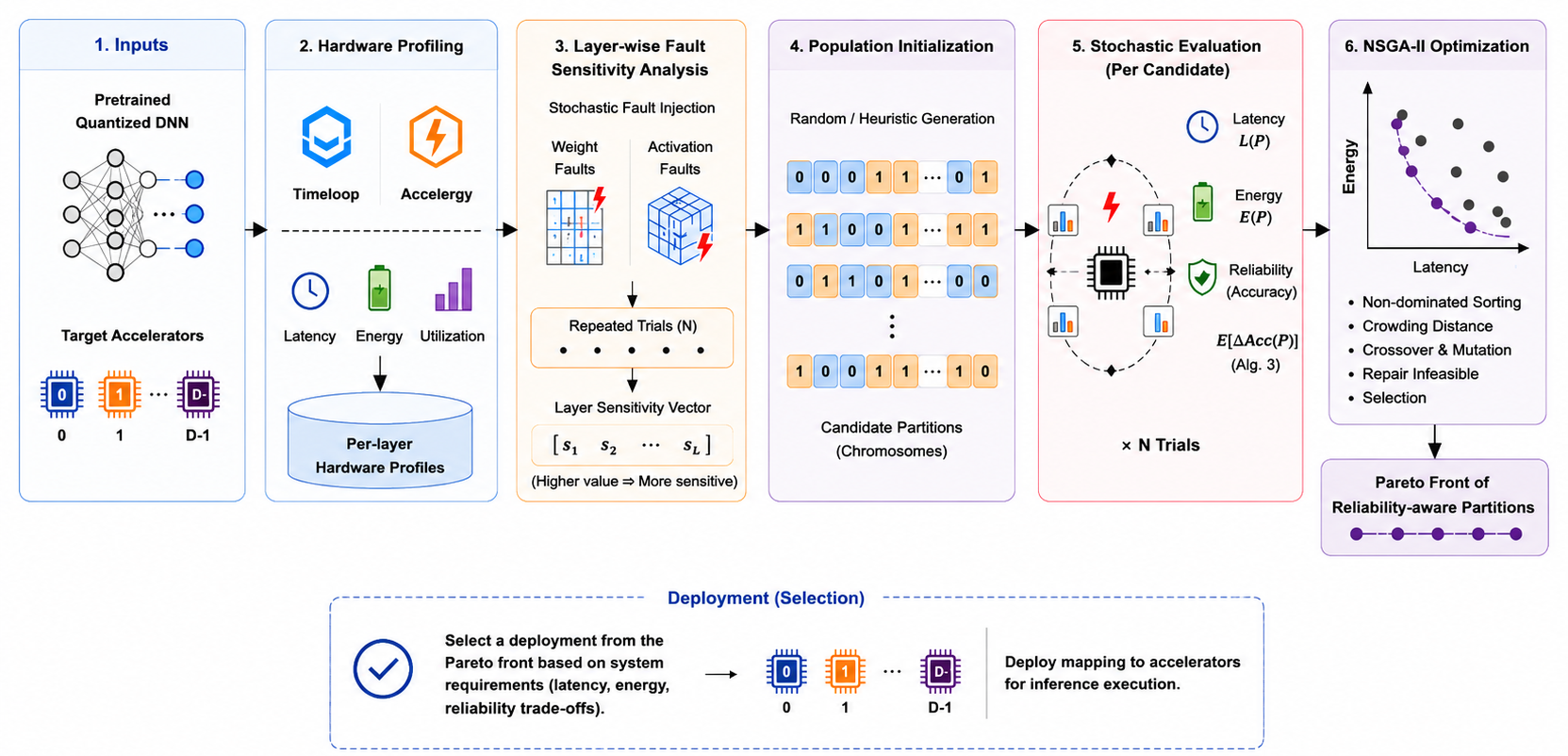} % Adjust width only
    % \vspace{-2mm} % Optional: reduce vertical space below the image
   \caption{The proposed RAPID-DNN framework. The framework combines hardware profiling, reliability-aware evaluation, and multi-objective optimization to derive Pareto-optimal DNN partitioning strategies that balance inference latency, energy consumption, and fault tolerance on heterogeneous edge accelerators.}
    % \vspace{-1mm} % Optional: reduce space below the caption
    \label{fig:framework}
\end{figure*}

% ============================================================

\section{Problem Formulation}
\label{sec:problem}

We formulate DNN partitioning as a three-objective optimization problem
that determines a layer-to-device mapping while jointly optimizing
inference latency, energy consumption, and fault resilience. Unlike
conventional partitioning approaches that assume fault-free execution,
the proposed formulation explicitly incorporates expected fault-induced
accuracy degradation as the third optimization objective.

%===========================================================

\subsection{Mathematical Framework}

Let a pretrained DNN consist of $L$ sequential layers deployed
across a heterogeneous edge platform comprising $D$ heterogeneous accelerators. A
partitioning configuration assigns every layer to exactly one
accelerator and is represented as

\begin{equation}
P:\{1,\ldots,L\}\rightarrow\{0,\ldots,D-1\},
\label{eq:partition}
\end{equation}

where $P(l)$ denotes the accelerator executing layer $l$.

Each accelerator is associated with layer-wise latency and energy
profiles (obtained from CNNParted), which are used to evaluate candidate
partitioning configurations.

To evaluate runtime reliability, RAPID-DNN employs bit-level
fault injection following the fault model described in
Section~\ref{sec:threat_model}. Runtime faults are independently
injected into the weight or activation tensors according to the
specified fault injection probability, and the resulting inference
accuracy is measured over repeated trials. Throughout this
work, the DNN topology and hardware performance profile are assumed to
remain fixed during each optimization cycle, while fault locations are
sampled independently across trials.

RAPID-DNN first characterizes the fault vulnerability of individual DNN
layers through layer-wise fault injection. The resulting vulnerability
profile is then used to evaluate candidate partitioning configurations
during optimization.
Let
$\mathrm{Acc}_{\mathrm{clean}}$ denote the fault-free inference
accuracy, and let $\mathrm{Acc}_k(l)$ denote the inference accuracy
obtained during the $k$-th trial when faults are injected
only into layer $l$, while all remaining layers execute without
perturbation. The corresponding layer-wise accuracy degradation is

\begin{equation}
\Delta\mathrm{Acc}_k(l)
=
\mathrm{Acc}_{\mathrm{clean}}
-
\mathrm{Acc}_k(l),
\label{eq:layer_drop}
\end{equation}

where fault locations within the selected layer are independently
sampled according to the fault injection probability. 

The resulting layer-wise sensitivity profile identifies
vulnerability-critical layers that are prioritized during
partition-level reliability evaluation. The sensitivity score $S_l$ is
estimated as

\begin{equation}
S_l
=
\frac{1}{N}
\sum_{k=1}^{N}
\Delta\mathrm{Acc}_k(l),
\label{eq:sensitivity}
\end{equation}

where $N$ denotes the number of independent fault-injection
trials. The collection

\[
\mathbf{S}
=
\{S_1,S_2,\ldots,S_L\}
\]

constitutes the layer-wise sensitivity profile of the network.

The resulting layer-wise sensitivity profile provides the basis for
evaluating the reliability of candidate partitioning configurations,
which is formulated as a multi-objective optimization problem in the
following subsection.

\subsection{Multi-Objective Optimization Strategy}

For a candidate partition configuration $P$, the first objective minimizes end-to-end inference latency,
\begin{equation}
\mathrm{Latency}(P)
=
\sum_{l=1}^{L}
t_l^{P(l)},
\label{eq:latency}
\end{equation}
where $t_l^{P(l)}$ denotes the execution latency of layer $l$ on the
accelerator assigned by partition $P$.

The second objective minimizes inference energy, given by
\begin{equation}
\mathrm{Energy}(P)
=
\sum_{l=1}^{L}
e_l^{P(l)},
\label{eq:energy}
\end{equation}
where $e_l^{P(l)}$ denotes the energy consumption of layer $l$ on the
accelerator assigned by partition $P$.

Each candidate partition is evaluated under
$N$ independent fault-injection trials to quantify reliability.
Let
$\mathrm{Acc}_{\mathrm{clean}}$
denote the fault-free inference accuracy and
$\mathrm{Acc}_k(P)$
denote the inference accuracy obtained during the
$k$-th fault-injection trial.
The corresponding fault-induced accuracy degradation is
\begin{equation}
\Delta\mathrm{Acc}_k(P)
=
\mathrm{Acc}_{\mathrm{clean}}
-
\mathrm{Acc}_k(P),
\label{eq:acc_drop}
\end{equation}
where fault locations are independently sampled for each trial
according to the fault injection probability defined in
Section~\ref{sec:threat_model}.

The third objective minimizes the expected degradation in inference accuracy caused by runtime hardware faults, and computed as
\begin{equation}
\mathbb{E}
\left[
\Delta\mathrm{Acc}(P)
\right]
=
\frac{1}{N}
\sum_{k=1}^{N}
\Delta\mathrm{Acc}_k(P).
\label{eq:exp_acc}
\end{equation}
% This serves as the reliability objective optimized by RAPID-DNN.
The expected fault-induced accuracy degradation serves as the
reliability objective, providing a quantitative measure of deployment
robustness under runtime hardware faults.
Unlike the layer-wise sensitivity score, which evaluates
individual layers independently, the
reliability objective measures
the expected accuracy degradation of the
entire partition configuration and therefore captures the
combined influence of hardware mapping, accelerator
characteristics and fault propagation.

Accordingly, RAPID-DNN formulates DNN partitioning as the following
three-objective optimization problem:
\begin{equation}
\min_{P}
\left(
\mathrm{Latency}(P),\;
\mathrm{Energy}(P),\;
\mathbb{E}
\left[
\Delta\mathrm{Acc}(P)
\right]
\right),
\label{eq:obj}
\end{equation}
subject to
\begin{align}
&
n_d(P)
\le
K_{\max},
\qquad
\forall d\le D,
\label{eq:contiguity}
\end{align}
where $n_d(P)$ denotes the number of contiguous layer segments assigned
to accelerator $d$, and $K_{\max}$ specifies the maximum number of
segments permitted per accelerator. The value of $K_{\max}$ depends on
the desired trade-off between partition flexibility and communication
overhead. A smaller $K_{\max}$ restricts partition fragmentation,
thereby limiting the number of inter-device communication boundaries,
whereas a larger value provides greater flexibility for
reliability-aware layer placement at the cost of additional
communication. In this work, the value of $K_{\max}$ is specified in the
experimental setup (Section~VI-A). The optimization objectives are
inherently conflicting because improving reliability may increase
inference latency and energy consumption.

% The number of contiguous layer segments assigned to accelerator $d$,
% denoted by $n_d(P)$, is equal to the number of contiguous layer blocks
% executed on accelerator $d$. Equivalently, it is obtained by counting
% the starting layer of each contiguous block.
A new segment begins whenever layer $l$ is assigned to accelerator $d$
and either $l=1$ or the preceding layer is assigned to a different
accelerator.
Formally,
\begin{equation}
n_d(P)
=
\left|
\left\{
l\in\{1,\ldots,L\}
:
P(l)=d,\;
\left(l=1\;\lor\;P(l-1)\neq d\right)
\right\}
\right|,
\label{eq:n_segments}
\end{equation}

where $|\cdot|$ denotes the cardinality of the set. 
% Thus, $n_d(P)$ counts the number of contiguous layer blocks assigned to
% accelerator $d$.
That is, $n_d(P)$ is obtained by counting the starting layer of each
contiguous segment assigned to accelerator $d$.
For example, if
$P=[0,0,1,1,0,2,2,0]$,
then
$n_0(P)=3$,
corresponding to the three contiguous layer segments
$(1,2)$,
$(5)$,
and
$(8)$.

% The optimization problem is combinatorial and the number of
% feasible layer-to-accelerator mappings grows exponentially with the
% number of DNN layers and available accelerators.
% Consequently,
% exhaustive search quickly becomes computationally infeasible for
% realistic networks. 
The search space grows exponentially with both the number of DNN layers and available accelerators, making exhaustive enumeration computationally impractical even for moderate-sized networks.
RAPID-DNN employs the NSGA-II multi-objective
evolutionary algorithm to efficiently approximate the Pareto front of
partitioning configurations that balance inference latency, energy
consumption and expected fault-induced accuracy degradation.

\section{Proposed Methodology}
\label{sec:method}

Figure~\ref{fig:framework} provides an overview of the proposed
RAPID-DNN workflow. The subsequent subsections describe the individual
components of the framework and their integration into the 
optimization process.

% ============================================================

\subsection{Reliability-Aware Partition Optimization}
\label{sec:optimization}

RAPID-DNN performs reliability-aware DNN partitioning through two
sequential stages, as illustrated in Fig.~\ref{fig:framework}:
(i) layer-wise reliability characterization and
(ii) reliability-aware multi-objective optimization.

Prior to optimization, hardware performance profiles are generated for
each accelerator using Timeloop and
Accelergy~\cite{timeloop,accelergy}. For every DNN layer, the profiling
process estimates latency and energy under the target precision. These
profiles are subsequently used to evaluate candidate partitions during
optimization. Table~\ref{tab:hw} summarizes the accelerator platforms
considered in this work. Eyeriss represents an energy-efficient spatial
dataflow architecture for CNN inference, whereas SIMBA represents a
high-performance multi-chip accelerator.

The first stage performs layer-wise reliability characterization by
independently evaluating each DNN layer through stochastic fault
injection. The resulting sensitivity profile identifies
vulnerability-critical layers that guide the proposed
sensitivity-guided evaluation strategy.
In the second stage, candidate partitions are encoded as chromosomes and
evaluated using the three-objective optimization formulation presented
in Section~\ref{sec:problem}. Latency and energy are obtained from the
precomputed hardware profiles, while the reliability objective is
estimated using the sensitivity-guided evaluation strategy. NSGA-II
iteratively evolves the population to obtain Pareto-optimal partitioning
solutions that jointly optimize inference latency, energy consumption,
and expected fault-induced accuracy degradation.

\begin{table}[htbp]
\caption{Heterogeneous Accelerator Profiles Used in RAPID-DNN}
\label{tab:hw}
\centering
\renewcommand{\arraystretch}{1.1}
\setlength{\tabcolsep}{2.5pt}

\begin{adjustbox}{max width=\textwidth}
\begin{tabular}{@{}lll@{}}
\toprule
\textbf{Property} &
\textbf{Eyeriss} &
\textbf{SIMBA} \\
\midrule

Architecture &
Spatial &
Multi-chip \\

Precision & INT16 (fixed point) & INT16 (fixed point) \\

Profiling &
Timeloop + Accelergy &
Timeloop + Accelergy \\

% Activity factor ($\alpha_d$) &
% 0.40 &
% 0.70 \\

\bottomrule
\end{tabular}
\end{adjustbox}

\vspace{1mm}
\end{table}

% \subsubsection{Layer-wise Reliability Characterization}
\subsection{Bit-Level Fault Injection}
\label{sec:fault_inject}

RAPID-DNN employs a unified stochastic bit-level fault-injection
framework to emulate transient runtime hardware faults during DNN
inference. The same fault-injection procedure is used for both
layer-wise reliability characterization and candidate partition
evaluation, ensuring a consistent fault model.

Algorithm~\ref{alg:inject} summarizes the stochastic bit-level
fault-injection procedure. Given a target weight or activation tensor,
individual elements are independently selected according to the fault
injection probability $p_f$. Selected elements are converted to an INT16
fixed-point representation, corrupted according to the specified fault
model, and then converted back to floating-point representation before
continuing inference. Algorithm~\ref{alg:inject} explicitly illustrates
the LSB, higher-order, sign-bit, and random fault models.
The corrupted tensor is subsequently used to evaluate the impact of
runtime hardware faults during DNN inference.

\noindent
For an input sample $x$, the inference process under a runtime hardware
fault injected at layer $l_{\mathrm{inj}}$ is represented as
\begin{equation}
\hat{y}
=
f_L
\circ
\cdots
\circ
f_{l_{\mathrm{inj}}+1}
\circ
\hat{f}_{l_{\mathrm{inj}}}
\circ
f_{l_{\mathrm{inj}}-1}
\circ
\cdots
\circ
f_1(x),
\label{eq:faulty_inf}
\end{equation}
where $\hat{y}$ denotes the fault-affected network output,
$f_l(\cdot)$ represents the computation performed by layer $l$, and
$\hat{f}_{l_{\mathrm{inj}}}$ denotes the computation of the
fault-injected layer after applying the specified bit-level fault. All
remaining layers execute normally on the resulting intermediate
activations.

A fault injected into layer $l_{\mathrm{inj}}$ may propagate through
subsequent layers and affect the final prediction. This fault-injection
process is used throughout RAPID-DNN for both layer-wise reliability
characterization and candidate partition evaluation.
% ============================================================

% ============================================================

% \begin{algorithm}[t]
% \caption{Stochastic Bit-Level Fault Injection}
% \label{alg:inject}
% \begin{algorithmic}[1]

% \REQUIRE Tensor $T$, fault probability $p_f$, fault mode $m$

% \ENSURE Corrupted tensor $T_f$

% \STATE $T_f \leftarrow T$

% \FORALL{element $x_i \in T_f$}

%     \IF{$\mathrm{Bernoulli}(p_f)=1$}

%         \IF{$m=$ LSB}
%             \STATE Flip one least significant bit
%         \ELSIF{$m=$ Higher-order}
%             \STATE Flip one higher-order magnitude bit
%         \ELSIF{$m=$ Sign-bit}
%             \STATE Flip the sign bit
%         \ENDIF

%     \ENDIF

% \ENDFOR

% \RETURN $T_f$

% \end{algorithmic}
% \end{algorithm}

\begin{algorithm}[t]
\caption{Stochastic Bit-Level Fault Injection}
\label{alg:inject}
\begin{algorithmic}[1]

\REQUIRE Floating-point tensor $T$, fault probability $p_f$, fault mode $m$
\ENSURE Corrupted tensor $T_f$

\STATE $T_f \leftarrow T$

\FORALL{element $x_i \in T_f$}

    \IF{$\mathrm{Bernoulli}(p_f)=1$}

        \STATE $q_i \leftarrow \mathrm{Quantize}(x_i)$ to INT16 fixed-point representation

        \IF{$m=$ LSB}
            \STATE Randomly select $b$ from $\{0,1,2,3\}$

        \ELSIF{$m=$ Higher-order}
            \STATE Randomly select $b$ from $\{10,\ldots,14\}$

        \ELSIF{$m=$ Sign-bit}
            \STATE Set $b \leftarrow 15$

        \ELSIF{$m=$ Random}
            \STATE Randomly select $b$ from $\{0,\ldots,15\}$

        \ENDIF

        \STATE $q_i \leftarrow \mathrm{FlipBit}(q_i, b)$

        \STATE $x_i \leftarrow \mathrm{Dequantize}(q_i)$

    \ENDIF

\ENDFOR

\RETURN $T_f$

\end{algorithmic}
\end{algorithm}

% ============================================================

%------------------------------------------------------------------
\subsection{Layer-Wise Reliability Characterization}
\label{sec:layerwise_method}
%------------------------------------------------------------------

RAPID-DNN first performs layer-wise reliability characterization to
quantify the vulnerability of individual DNN layers before partition
optimization. Rather than assuming uniform layer-wise fault sensitivity, the
proposed framework identifies the vulnerability-critical layers whose
corruption contributes most significantly to inference accuracy
degradation. 
The sensitivity profile obtained is used only to identify the vulnerability-critical layer set that guides candidate partition evaluation.

\textbf{Layer-wise sensitivity profiling.}
Each DNN layer is independently characterized by injecting runtime
faults only into that layer while all remaining layers execute without
perturbation. Repeated independent fault-injection trials are performed,
and the resulting fault-induced accuracy degradation is averaged to
estimate the layer sensitivity score according to the formulation in
Section~\ref{sec:problem}. The resulting sensitivity profile is then
ranked in descending order of layer sensitivity. 
A fraction $\rho$ of the highest-ranked layers is selected as the
vulnerability-critical layer set. 
The number of selected critical layers is computed as
\[
K=\min\!\left(3,\;\max\!\left(2,\;\mathrm{round}(\rho L)\right)\right),
\]
where $L$ denotes the total number of network layers. 
This keeps the guided fault-injection process compact regardless of network depth: the lower bound of 2 rules out the degenerate case of a single critical layer, while the upper bound of 3 keeps candidate evaluation during NSGA-II optimization computationally manageable. 
The same
criterion is applied consistently across all DNN models. With
$\rho=0.15$, it selects two critical layers for shallow networks such as
AlexNet ($L=8$) and three critical layers for deeper networks such as
ResNet18 ($L=20$) and MobileNetV2 ($L=52$), providing a compact
critical-layer set across networks with different depths.

Algorithm~\ref{alg:layerwise} summarizes the layer-wise reliability
characterization procedure. The same fault-injection framework, fault
models, and injection probabilities described in
Section~\ref{sec:fault_inject} are employed throughout the profiling
process. 
% Unless otherwise stated, each layer is evaluated using
% $N=50$ independent fault-injection trials, and the top $\rho L$ layers
% are selected as the vulnerability-critical layer set.

% \begin{algorithm}[t]
% \caption{Layer-Wise Reliability Characterization}
% \label{alg:layerwise}
% \begin{algorithmic}[1]

% \REQUIRE DNN model $M$, dataset $\mathcal{D}$, fault configuration $\mathcal{F}$, number of trials $N$, critical-layer selection ratio $\rho$

% \ENSURE Layer-wise sensitivity profile $\mathbf{S}$, vulnerability-critical layer set $\mathcal{C}$

% \STATE Compute the fault-free accuracy $\mathrm{Acc}_{\mathrm{clean}}$

% \FOR{$l=1$ \textbf{to} $L$}

%     \STATE $\Delta_l \leftarrow 0$

%     \FOR{$k=1$ \textbf{to} $N$}

%         \STATE Apply stochastic fault injection only to layer $l$
%         using Algorithm~\ref{alg:inject}

%         \STATE Execute inference and obtain
%         $\mathrm{Acc}_{k}^{(l)}$

%         \STATE
%         $\Delta_l \leftarrow
%         \Delta_l+
%         \left(
%         \mathrm{Acc}_{\mathrm{clean}}
%         -
%         \mathrm{Acc}_{k}^{(l)}
%         \right)$

%     \ENDFOR

%     \STATE
%     $S_l \leftarrow \Delta_l/N$

% \ENDFOR

% \STATE
% Construct the sensitivity profile
% $\mathbf{S}=[S_1,S_2,\ldots,S_L]$

% \STATE
% Rank the layers in descending order of $S_l$

% \STATE
% $K=\min(3,\max(2,\mathrm{round}(\rho L)))$
% \STATE
% Select the highest-ranked $K$ layers to form the
% vulnerability-critical layer set $C$

% \RETURN $\mathbf{S},\mathcal{C}$

% \end{algorithmic}
% \end{algorithm}

\begin{algorithm}[t]
\caption{Layer-Wise Reliability Characterization}
\label{alg:layerwise}
\begin{algorithmic}[1]

\REQUIRE DNN model $M$ with $L$ layers, dataset $\mathcal{D}$,
fault configuration $\mathcal{F}$, number of trials $N$,
critical-layer selection ratio $\rho$

\ENSURE Layer-wise sensitivity profile $\mathbf{S}$,
vulnerability-critical layer set $\mathcal{C}$

\STATE Compute the fault-free inference accuracy
$\mathrm{Acc}_{\mathrm{clean}}$

\FOR{$l=1$ \textbf{to} $L$}

    \STATE $\Delta_l \leftarrow 0$

    \FOR{$k=1$ \textbf{to} $N$}

        \STATE Apply stochastic fault injection only to layer $l$
        using Algorithm~\ref{alg:inject}

        \STATE Execute inference and obtain
        $\mathrm{Acc}_{k}^{(l)}$

        \STATE
        $\Delta_l \leftarrow
        \Delta_l +
        \left(
        \mathrm{Acc}_{\mathrm{clean}}
        -
        \mathrm{Acc}_{k}^{(l)}
        \right)$

    \ENDFOR

    \STATE
    $S_l \leftarrow \Delta_l/N$

\ENDFOR

\STATE
Construct the sensitivity profile
$\mathbf{S}=[S_1,S_2,\ldots,S_L]$

\STATE
Rank the layers in descending order of $S_l$;
break ties by ascending layer index

\STATE
$K \leftarrow
\min(3,\max(2,\mathrm{round}(\rho L)))$

\STATE
Select the first $K$ ranked layers to form the
vulnerability-critical layer set $\mathcal{C}$

\RETURN $\mathbf{S},\mathcal{C}$

\end{algorithmic}
\end{algorithm}

\textbf{Vulnerability-critical layer selection.}
The ranked sensitivity profile is used to construct the
vulnerability-critical layer set
\[
\mathcal C
=
\{l_1,l_2,\ldots,l_K\},
\]
which is supplied to the candidate partition evaluation stage.
Prioritizing these layers focuses the stochastic fault-injection
evaluation on the identified vulnerability-critical layers while still
assessing the reliability of the complete partition configuration.
The layer-wise sensitivity scores are used only to identify
vulnerability-critical layers for guided fault-injection sampling. The
reliability objective is subsequently computed as the expected
fault-induced accuracy degradation of the complete candidate partition.

\subsection{Candidate Partition Representation and Evaluation}
\label{subsec:partition_eval}

\begin{figure}[t!]
    \centering
    \includegraphics[width=\linewidth]{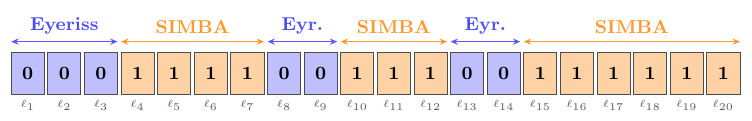} % Adjust width only
    % \vspace{-2mm} % Optional: reduce vertical space below the image
     \caption{Example chromosome encoding for ResNet18
($L=20$, $D=2$, $K_{\max}=3$). Each gene specifies the accelerator
assignment for a network layer ($0=$ Eyeriss, $1=$ SIMBA). The
illustrated chromosome comprises six contiguous execution segments
($n_0=n_1=3$), satisfying the contiguity constraint defined in
Eq.~\eqref{eq:contiguity}.}
    % \vspace{-1mm} % Optional: reduce space below the caption
    \label{fig:chromosome}
\end{figure}

During NSGA-II optimization, each candidate partition is represented as
a chromosome of length $L$,

\begin{equation}
\mathbf{c}
=
[d_1,d_2,\ldots,d_L],
\qquad
d_l\in\{0,1,\ldots,D-1\},
\end{equation}

where gene $d_l$ specifies the accelerator assigned to execute layer
$l$. Thus, each chromosome uniquely represents a complete
layer-to-device mapping for the DNN. Figure~\ref{fig:chromosome}
illustrates an example chromosome for a heterogeneous
two-accelerator deployment.

% ============================================================

Each candidate chromosome is evaluated using the precomputed hardware
performance profiles to estimate inference latency and energy
consumption, while reliability is obtained through stochastic fault
injection over the complete partitioned DNN. Unlike the layer-wise
characterization stage, which evaluates individual layers, this stage
assesses the reliability of the deployed network under the
chromosome-defined layer-to-accelerator mapping.

The vulnerability-critical layer set $\mathcal{C}$ identified during
layer-wise profiling guides the reliability evaluation. These layers contribute most to fault-induced accuracy degradation and
are therefore sampled more frequently, while additional random trials on
the remaining layers ensure that the reliability estimate reflects the
behavior of the entire network.

The reliability objective is defined as the average fault-induced
accuracy degradation over all guided and random fault-injection trials.
Candidate partitions that reduce the impact of faults on
vulnerability-critical layers therefore incur lower reliability costs
while still accounting for the influence of the remaining layers.
Candidate chromosomes violating the contiguity constraint are discarded
before fitness evaluation.
Algorithm~\ref{alg:robust} summarizes the evaluation procedure. The
guided evaluation performs a total of
$N_c=|\mathcal{C}|\times n_c$ fault-injection trials, where each
vulnerability-critical layer is evaluated independently over $n_c$
trials. This is followed by $N_r$ additional random trials on the
remaining layers to estimate the reliability of the complete
partitioned DNN.

The guided fault-injection procedure is used during
NSGA-II optimization to estimate the reliability objective of candidate
partitions. A reduced evaluation budget is employed for computational
efficiency and is used only for fitness estimation, whereas the final
experimental results are obtained using the full stochastic
fault-injection protocol.

\begin{algorithm}[t]
\caption{Candidate Partition Evaluation}
\label{alg:robust}
\begin{algorithmic}[1]

\REQUIRE
Candidate partition $P$,
hardware profile $\mathcal{H}$,
critical-layer set $\mathcal{C}$,
fault configuration $\mathcal{F}$,
$n_c$ guided trials per critical layer,
$N_r$ random trials

\ENSURE
$\mathrm{Latency}(P)$,
$\mathrm{Energy}(P)$,
$\mathbb{E}[\Delta\mathrm{Acc}(P)]$

\STATE Estimate $\mathrm{Latency}(P)$ using $\mathcal{H}$

\STATE Estimate $\mathrm{Energy}(P)$ using $\mathcal{H}$

\STATE Compute fault-free accuracy $\mathrm{Acc}_{\rm clean}$

\STATE $\Delta_{\rm sum}\leftarrow0$

\FORALL{$l\in\mathcal{C}$}

    \FOR{$k=1$ \textbf{to} $n_c$}

        \STATE Apply stochastic fault injection to layer $l$
        using Algorithm~\ref{alg:inject}

        \STATE Execute partitioned inference using partition $P$

        \STATE Measure $\mathrm{Acc}(P)$

        \STATE
        $\Delta_{\rm sum}
        \leftarrow
        \Delta_{\rm sum}
        +
        \left(
        \mathrm{Acc}_{\rm clean}
        -
        \mathrm{Acc}(P)
        \right)$

    \ENDFOR

\ENDFOR

\FOR{$k=1$ \textbf{to} $N_r$}

    \STATE Apply stochastic fault injection simultaneously to
    all non-critical layers
    ${\{1,\ldots,L\}\setminus\mathcal{C}}$
    using Algorithm~\ref{alg:inject}

    \STATE Execute partitioned inference using partition $P$

    \STATE Measure $\mathrm{Acc}(P)$

    \STATE
    $\Delta_{\rm sum}
    \leftarrow
    \Delta_{\rm sum}
    +
    \left(
    \mathrm{Acc}_{\rm clean}
    -
    \mathrm{Acc}(P)
    \right)$

\ENDFOR

\STATE
$\mathbb{E}[\Delta\mathrm{Acc}(P)]
\leftarrow
\Delta_{\rm sum}/(|\mathcal{C}|\,n_c+N_r)$

\RETURN
$\mathrm{Latency}(P)$,
$\mathrm{Energy}(P)$,
$\mathbb{E}[\Delta\mathrm{Acc}(P)]$

\end{algorithmic}
\end{algorithm}

\subsection{Fault-Aware Multi-Objective Partition Optimization}
\label{subsec:optimization}

RAPID-DNN formulates DNN partitioning as a three-objective optimization
problem that simultaneously minimizes inference latency, energy
consumption, and expected fault-induced accuracy degradation. The
optimization is performed using the NSGA-II evolutionary algorithm,
which searches the space of feasible layer-to-device mappings while
maintaining a diverse set of non-dominated solutions.

% Algorithm~\ref{alg:opt} summarizes the optimization procedure. An
% initial population of feasible candidate partitions is first generated
% and evaluated using the latency, energy, and reliability objectives.
% Latency and energy are obtained directly from the precomputed hardware
% performance profiles, whereas reliability is estimated using the guided
% stochastic fault-injection procedure described in
% Algorithm~\ref{alg:robust}. During each generation, offspring are
% produced through selection, crossover, and mutation. Since these
% genetic operations may violate the maximum contiguous-segment
% constraint, a repair operator is applied before fitness evaluation. The
% repair procedure iteratively merges violating contiguous layer segments
% until the contiguity constraint is satisfied, ensuring that all
% offspring evaluated by NSGA-II correspond to feasible deployment
% configurations.
% The repaired offspring are subsequently evaluated and combined with the
% parent population. NSGA-II then performs non-dominated sorting and
% crowding-distance assignment, followed by elitist selection to form the
% next generation. The optimization converges to a Pareto front
% representing different trade-offs among inference latency, energy
% consumption, and runtime reliability. The resulting Pareto-optimal
% partitions provide multiple deployment strategies, allowing designers
% to select the solution that best satisfies the latency, energy, and
% reliability requirements of the target FPGA platform.

Algorithm~\ref{alg:opt} summarizes the optimization procedure. The
initial population is generated by randomly assigning each DNN layer to
one of the available accelerators, producing candidate layer-to-device
mappings. Each feasible candidate partition is then evaluated using the
latency, energy, and reliability objectives.
Latency and energy are obtained from the precomputed hardware
performance profiles, whereas reliability is estimated using the guided
stochastic fault-injection procedure described in
Algorithm~\ref{alg:robust}. During each generation, offspring are
generated using two-point crossover with probability $p_c$ and
uniform integer mutation with probability $p_m$.
Since these operations may violate the maximum contiguous-segment
constraint, a repair operator is applied before fitness evaluation. The
repair procedure merges violating contiguous layer segments until the
constraint is satisfied, ensuring that only feasible partition
configurations are evaluated.
The repaired offspring are then evaluated and merged with the parent
population. NSGA-II performs non-dominated sorting on the
three-objective fitness values, computes the crowding distance, and
applies elitist environmental selection to construct the next
generation. The optimization converges to a Pareto front representing
the trade-offs among inference latency, energy consumption, and
expected fault-induced accuracy degradation. The resulting Pareto-optimal
partitions provide alternative deployment strategies, allowing
designers to select a solution that best meets the performance and
reliability requirements of the target deployment.

% \begin{algorithm}[t]
% \caption{Fault-Aware Multi-Objective Partition Optimization}
% \label{alg:opt}
% \begin{algorithmic}[1]

% \REQUIRE
% DNN model $M$,
% hardware profile $\mathcal{H}$,
% critical-layer set $\mathcal{C}$,
% population size $N_{\rm pop}$,
% number of generations $G$

% \ENSURE
% Pareto-optimal partition set $\mathcal{P}^{*}$

% \STATE Generate an initial population of feasible layer-to-device mappings

% \STATE Evaluate all candidate partitions using Algorithm~\ref{alg:robust}

% \FOR{$g=1$ \textbf{to} $G$}

%     \STATE Generate offspring through selection, crossover, and mutation

%     \STATE Repair offspring by enforcing the contiguity constraint

%     \STATE Evaluate repaired offspring using Algorithm~\ref{alg:robust}

%     \STATE Combine parent and offspring populations

%     \STATE Perform non-dominated sorting

%     \STATE Compute crowding distance

%     \STATE Perform elitist selection to create the next generation

% \ENDFOR

% \RETURN Pareto-optimal partition set $\mathcal{P}^{*}$

% \end{algorithmic}
% \end{algorithm}

\begin{algorithm}[t]
\caption{Fault-Aware Multi-Objective Partition Optimization}
\label{alg:opt}
\begin{algorithmic}[1]

\REQUIRE
DNN model $M$,
hardware profile $\mathcal{H}$,
critical-layer set $\mathcal{C}$,
population size $N_{\rm pop}$,
number of generations $G$,
crossover probability $p_c$,
mutation probability $p_m$

\ENSURE
Pareto-optimal partition set $\mathcal{P}^{*}$

\STATE Generate an initial population of feasible layer-to-device mappings

\STATE Evaluate each candidate partition using
Algorithm~\ref{alg:robust} to obtain the objective vector
$(\mathrm{Latency}(P),\,
\mathrm{Energy}(P),\,
\mathbb{E}[\Delta\mathrm{Acc}(P)])$

\FOR{$g=1$ \textbf{to} $G$}

    \STATE Generate offspring by applying two-point crossover
    to consecutive parent pairs with probability $p_c$, followed by
    uniform integer mutation with probability $p_m$

    \STATE Repair infeasible offspring by enforcing the contiguity constraint

    \STATE Evaluate repaired offspring using Algorithm~\ref{alg:robust}

    \STATE Combine parent and offspring populations

    \STATE Perform non-dominated sorting based on the objective vectors

    \STATE Compute crowding distance

    \STATE Select the top $N_{\rm pop}$ individuals from the combined
    population according to non-domination rank and crowding distance
    to form the next generation

\ENDFOR

\RETURN Pareto-optimal partition set $\mathcal{P}^{*}$

\end{algorithmic}
\end{algorithm}

%===========================================================
\section{Experimental Evaluation}
\label{sec:experiments}
%===========================================================

This section evaluates the effectiveness of the proposed RAPID-DNN
framework using five representative CNN models deployed on
heterogeneous edge accelerator profiles under transient runtime
hardware faults. The evaluation examines the proposed layer-wise
reliability characterization, assesses the effectiveness of
reliability-aware partition optimization, analyzes robustness under
different fault models and injection probabilities, and quantifies the
associated latency--energy trade-offs.

The experimental evaluation is designed to comprehensively assess the
effectiveness of RAPID-DNN from five complementary perspectives. First,
we examine the ability of the proposed layer-wise reliability
characterization to accurately identify vulnerability-critical DNN
layers. Next, we evaluate whether RAPID-DNN improves inference
robustness compared with fault-agnostic DNN partitioning under runtime
hardware faults. We then investigate the robustness of the proposed
framework under varying fault injection probabilities and bit-level
fault models. Subsequently, we analyze the latency and energy overheads
incurred in achieving improved runtime reliability. Finally, an ablation
study is conducted to quantify the contribution of each major component
of RAPID-DNN to the overall reliability improvement.

%===========================================================

%===========================================================

\subsection{Setup}
\label{sec:setup}

Table~\ref{tab:exp_setup} summarizes the experimental configuration.
RAPID-DNN is implemented in Python using PyTorch for DNN inference and
the \texttt{DEAP} library for NSGA-II optimization. Per-layer latency
and energy profiles are generated using Timeloop and
Accelergy~\cite{timeloop,accelergy} following the profiling methodology
adopted by CNNParted. Latency and energy for each partition are obtained
by aggregating the corresponding layer profiles. All DNN models are
imported in ONNX format and executed using the PyTorch inference
framework.
% Experiments are conducted using a fixed evaluation subset of 512 images
% from the publicly available ImageNet-10K
% dataset~\cite{rana2023imagenet10k}, derived from the official
% ILSVRC2012 ImageNet validation set~\cite{imagenet}. The subset is
% constructed using a deterministic sampling procedure with a fixed random
% seed (42), selecting images in a round-robin manner across the
% 1,000 ImageNet classes to ensure broad class coverage and
% reproducibility. Consequently, the clean Top-1 accuracies reported in
% Table~\ref{tab:dnn_summary} correspond to this subset and should not be
% directly compared with the standard ImageNet-1k benchmark results
% reported on the complete 50,000-image ILSVRC2012 validation set. The
% same evaluation subset is used for all fault-injection experiments and
% for both RAPID-DNN and CNNParted.
Experiments are conducted using a fixed evaluation subset of 512 images
from the publicly available ImageNet-10K
dataset~\cite{rana2023imagenet10k}, derived from the official
ILSVRC2012 ImageNet validation set~\cite{imagenet}. The same fixed
evaluation subset is used consistently for both RAPID-DNN and
CNNParted throughout all experiments to ensure a fair comparative
evaluation.
Two representative accelerator architectures are considered: Eyeriss,
an energy-efficient spatial DNN accelerator, and SIMBA, a
high-performance multi-chip accelerator.

Fault injection follows the threat model described in
Section~\ref{sec:threat_model}. Runtime faults are injected into the
weight and activation domains using random, LSB, higher-order, and
sign-bit fault models under multiple fault injection probabilities. The
corresponding parameters are summarized in
Table~\ref{tab:fault_setup}. Fault locations
are sampled independently according to the specified fault probability.
Partition optimization is performed using NSGA-II with a population
size of 60 and 60 generations. 
The crossover and mutation probabilities were fixed to
$p_c=0.7$ and $p_m=0.3$, respectively, and were used
consistently throughout all experiments.
The optimization jointly considers the
three objectives defined in Section~\ref{sec:problem}: expected
fault-induced accuracy degradation, end-to-end inference latency, and
inference energy consumption. The critical-layer
selection ratio is fixed at $\rho=0.15$, with
$K=\min(3,\max(2,\mathrm{round}(\rho L)))$. The maximum number of
partition segments is fixed at $K_{\max}=3$ for all evaluated DNN
architectures and accelerator configurations.

All reported reliability results are averaged
over $N=50$ independent whole-network fault-injection trials and are
reported with 95\% confidence intervals. This evaluation protocol is
used only for the final experimental results and is distinct from the
guided fault-injection procedure employed during NSGA-II optimization.
Preliminary convergence analysis showed that both the estimated
fault-induced accuracy degradation and its 95\% confidence interval
stabilized after approximately 40--50 trials; therefore, $N=50$ was
adopted throughout the experiments to balance statistical confidence and
computational cost.

RAPID-DNN is compared with CNNParted, a multi-objective DNN
partitioning framework that optimizes inference latency and energy
consumption without explicitly considering runtime reliability. Both
frameworks use identical hardware profiles, datasets, fault-injection
settings, and optimization parameters. An ablation baseline is obtained
by disabling the proposed reliability objective while retaining the
remaining optimization pipeline.
%===========================================================

\begin{table}[htbp]
\caption{Experimental Configuration}
\label{tab:exp_setup}
\centering
\footnotesize
\renewcommand{\arraystretch}{1.2}
\begin{tabularx}{\columnwidth}{@{}lX@{}}
\toprule
\textbf{Category} & \textbf{Configuration} \\
\midrule
Implementation & Python + PyTorch \\
Optimization & NSGA-II (DEAP) \\
Hardware profiling & Timeloop + Accelergy \\
Accelerators & Eyeriss, SIMBA \\
DNN models & AlexNet, SqueezeNet, ResNet18, VGG16, MobileNetV2 \\
Inference precision & 16-bit fixed point \\
Optimization objectives & Latency, Energy, Expected accuracy degradation \\
Fault trials & $N=50$ \\
Population size & 60 \\
Generations & 60 \\
Crossover operator & Two-point crossover \\
Crossover probability ($p_c$) & 0.7 \\
Mutation operator & Uniform integer mutation \\
Mutation probability ($p_m$) & 0.3 \\
Baseline & CNNParted \\
Critical-layer ratio ($\rho$) & 0.15 \\
Maximum number of segments ($K_{\max}$) & 3 \\
\bottomrule
\end{tabularx}
\end{table}

%===========================================================

\begin{table}[htbp]
\caption{Fault Injection Parameters}
\label{tab:fault_setup}
\centering
\footnotesize
\renewcommand{\arraystretch}{1.2}
\begin{adjustbox}{max width=\textwidth}
\begin{tabular}{ll}
\toprule
\textbf{Parameter} & \textbf{Configuration} \\
\midrule
Fault domains &
Weights, Activations \\

Fault models &
Random, LSB, Higher-order, Sign-bit \\

Fault injection probabilities &
0.1\%, 0.5\%, 1\%, 2\%, 5\% \\

Stress-test scenarios &
10\%, 20\% \\

Injection granularity &
Bit-level perturbation \\

Evaluation method &
Stochastic fault injection \\

Sampling strategy&
Independent Bernoulli trials \\

Accuracy metric &
Top-1 classification accuracy \\

Reliability metric &
Expected accuracy degradation \\

Statistical analysis &
Mean over $N=50$ trials with \\
&
95\% confidence intervals \\

\bottomrule
\end{tabular}
\end{adjustbox}
\end{table}

\begin{table}[htbp]
\caption{Characteristics of the evaluated DNN benchmarks. Top-1 accuracy
is reported on the fixed 512-image evaluation subset derived from the
ImageNet-10K dataset.}
\label{tab:dnn_summary}
\centering
\footnotesize
\renewcommand{\arraystretch}{1.15}
\setlength{\tabcolsep}{4pt}
\begin{adjustbox}{max width=\textwidth}
\begin{tabular}{lcccc}
\toprule
\textbf{Model} &
\textbf{Layers} &
\textbf{Parameters} &
\textbf{Top-1 (\%)} &
\textbf{FLOPs} \\
\midrule
AlexNet      & 8  & 61.10 M  & 77.1 & 0.71 G \\
SqueezeNet   & 18 & 1.24 M  & 67.4 & 0.35 G \\
ResNet18     & 20 & 11.69 M  & 85.2 & 1.81 G \\
VGG16        & 16 & 138.36 M & 87.9 & 15.47 G \\
MobileNetV2  & 52 & 3.50 M  & 84.4 & 0.30 G \\
\bottomrule
\end{tabular}
\end{adjustbox}
\end{table}

%===========================================================

%===========================================================
\subsection{Layer-Wise Reliability Evaluation}
\label{sec:layer_results}
%===========================================================

This subsection evaluates the proposed layer-wise reliability
characterization by analyzing the vulnerability of individual DNN layers
to runtime hardware faults and assessing its effectiveness in
identifying vulnerability-critical layers.
Each DNN layer is evaluated
independently by injecting transient runtime faults while all
remaining layers execute fault-free. The average Top-1 accuracy
degradation over multiple stochastic fault-injection trials is used to
compute the layer sensitivity score defined in
Section~\ref{sec:method}. Layers exhibiting the highest sensitivity
are selected as the vulnerability-critical layer set
$\mathcal{C}$, which subsequently guides the reliability-aware
partition optimization.

\begin{figure*}[t]
\centering
\includegraphics[width=0.95\textwidth]{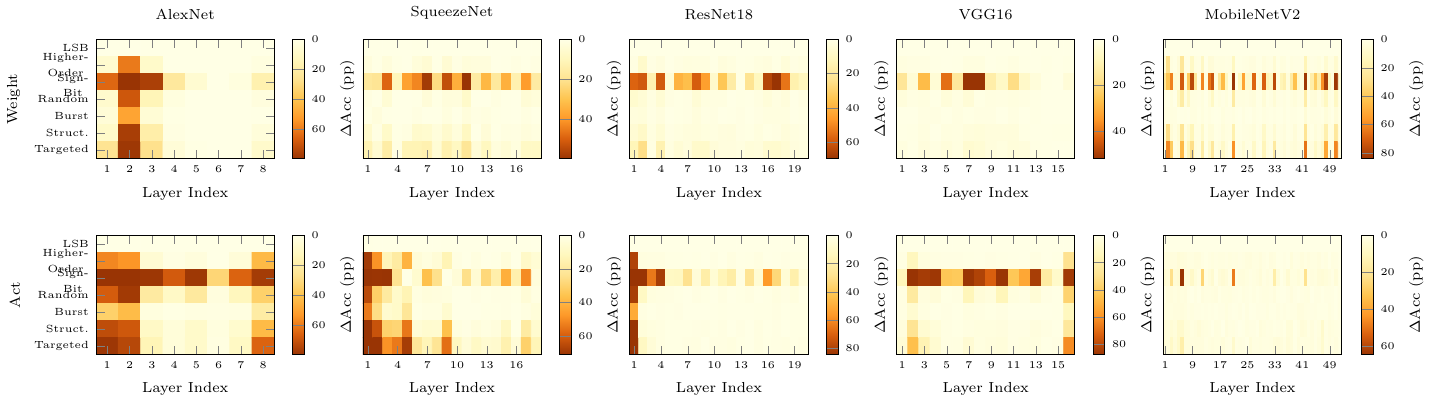}
\caption{Layer-wise reliability characterization of the evaluated DNN
architectures under LSB, higher-order, sign-bit, random, burst,
structured, and targeted fault injection. Each cell represents the
average Top-1 accuracy degradation of the corresponding layer at a fixed
fault injection probability ($p_f = 10^{-2}$). Darker colors indicate
greater layer vulnerability. Each heatmap uses an independent color
scale; therefore, color intensities should be compared only within the
same DNN architecture.}
\label{fig:layer_heatmaps}
\end{figure*}

%===========================================================

\subsubsection{Layer-Wise Reliability Analysis}

Figure~\ref{fig:layer_heatmaps} presents the resulting layer-wise
sensitivity profiles for the five evaluated CNN architectures under
seven fault modes, namely LSB, higher-order, sign-bit, random, burst,
structured, and targeted fault injection. 
Each heatmap reports the average Top-1 accuracy degradation obtained from isolated fault injection into individual layers at a fixed fault injection probability of 1\%. For each layer, the reported sensitivity is averaged over $N=50$ independent fault-injection trials, consistent with the experimental protocol described in Section~\ref{sec:setup}.
Although runtime faults
affect every network, the resulting accuracy degradation is highly
non-uniform. Only a small subset of layers produces substantial
degradation, whereas the remaining layers exhibit comparatively low
sensitivity. Moreover, the locations of these highly vulnerable layers
vary across architectures, indicating that fault sensitivity depends on
both the network topology and the computational characteristics of the
layers rather than on layer depth alone.
For visualization purposes, each heatmap is independently normalized
using the minimum and maximum layer sensitivity observed within the
corresponding network. Consequently, darker colors indicate relatively
more vulnerable layers within the same architecture, whereas lighter
colors indicate comparatively less sensitive layers. Since each network
uses its own color scale, color intensities should not be compared
directly across different DNN architectures.

\begin{table*}[htbp]
\caption{Vulnerability-critical layers identified from the layer-wise
reliability characterization shown in
Fig.~\ref{fig:layer_heatmaps}. The sensitivity score $S_l$ is computed
by averaging the per-layer Top-1 accuracy degradation across all
evaluated fault modes and fault domains. The reported layers are the
highest-ranked according to $S_l$, with the number of selected layers
determined by the critical-layer ratio $\rho$
(Algorithm~\ref{alg:layerwise}).}
\label{tab:critical_identification}
\centering
\footnotesize
\begin{adjustbox}{max width=\textwidth}
\begin{tabular}{lccc}
\toprule
\textbf{Network} & \textbf{Critical Layer} & \textbf{Layer Type} & \textbf{Sensitivity Score $S_l$ (pp)} \\
\midrule
AlexNet     & Conv2                     & Convolution                  & 57.85 \\
            & Conv1                     & Convolution (stem)           & 36.35 \\
\midrule
SqueezeNet  & Conv1                     & Convolution (stem)           & 31.53 \\
            & Fire2 (squeeze)           & Fire Module                  & 24.05 \\
            & Fire2 (expand)            & Fire Module                  & 19.01 \\
\midrule
ResNet18    & Conv1                     & Convolution (stem)           & 39.25 \\
            & Layer1.0--Conv1           & Residual Block               & 15.89 \\
            & Layer1.1--Conv1           & Residual Block               & 12.70 \\
\midrule
VGG16       & FC8                       & Fully Connected (classifier) & 18.25 \\
            & Conv1\_2                  & Convolution                  & 15.14 \\
\midrule
MobileNetV2 & Bottleneck 2 (pointwise)  & Inverted Residual            & 20.37 \\
            & Bottleneck 7 (pointwise)  & Inverted Residual            & 18.36 \\
            & Bottleneck 14 (pointwise) & Inverted Residual            & 17.63 \\
\bottomrule
\end{tabular}
\end{adjustbox}
\end{table*}

\subsubsection{Identification of Vulnerability-Critical Layers}

~\autoref{tab:critical_identification} summarizes the
vulnerability-critical layers identified from the aggregated
layer sensitivity score $S_l$. The results indicate that only a
small fraction of layers dominates the fault sensitivity of each
network. For example, AlexNet exhibits the highest sensitivity in
Conv2 ($S_l=57.85$ pp), followed by Conv1 ($36.35$ pp), whereas
SqueezeNet is primarily influenced by the stem convolution
($31.53$ pp) and the Fire2 module (24.05 pp and 19.01 pp). In
ResNet18, the stem convolution remains the most vulnerable layer
($39.25$ pp), with substantially lower sensitivities observed for
the early residual blocks (15.89 and 12.70 pp). In contrast,
VGG16 exhibits its highest sensitivity in the classifier layer
FC8 ($18.25$ pp), while MobileNetV2 is dominated by intermediate
pointwise convolutions within the inverted residual blocks
(17.63--20.37 pp). The variation in both the locations and
magnitudes of the highest-ranked layers across architectures
demonstrates that runtime fault sensitivity is strongly
architecture dependent, thereby motivating the use of
application-specific layer-wise reliability characterization
rather than a fixed heuristic for identifying critical layers.
The identified vulnerability-critical layers constitute only a
small fraction of the overall network depth, yet they exhibit
significantly higher fault sensitivity than the remaining layers. Consequently, RAPID-DNN focuses the
subsequent reliability evaluation on these layers, enabling the
optimizer to use the layer sensitivity scores obtained from the
preceding reliability characterization rather than assuming
uniform vulnerability across all layers.

\begin{table*}[htbp]
\caption{Reliability improvement achieved by RAPID-DNN using
the vulnerability-critical layer set identified through the proposed
layer-wise reliability characterization. $|\mathcal{C}|$ denotes the
number of vulnerability-critical layers selected for candidate
partition evaluation, while ``Reassigned'' indicates the number of
critical layers assigned to different accelerators compared with the
CNNParted solution. 
Accuracy values correspond to Top-1 inference accuracy under the random
bit-flip fault model at a 10\% fault injection probability (the
configuration of Table~\ref{tab:fault_setup}).}
\label{tab:critical_layers}

\centering
\footnotesize
\renewcommand{\arraystretch}{1.15}
\setlength{\tabcolsep}{5pt}

\begin{adjustbox}{max width=\textwidth}
\begin{tabular}{lcccccc}

\toprule

\textbf{Network}
&
$|\mathcal{C}|$
&
\textbf{Critical Layers}
&
\textbf{CNNParted}
&
\textbf{RAPID-DNN}
&
\textbf{Gain}
&
\textbf{Reduction in}
\\

&
&
\textbf{Reassigned}
&
\textbf{Top-1 (\%)}
&
\textbf{Top-1 (\%)}
&
\textbf{(pp)}
&
\textbf{Accuracy Degradation (\%)}

\\

\midrule

AlexNet
&
2
&
1
&
13.54
&
\textbf{17.15}
&
+3.61
&
5.7

\\

SqueezeNet
&
3
&
2
&
23.57
&
\textbf{42.06}
&
+18.5
&
42.2

\\

ResNet18
&
3
&
2
&
63.98
&
\textbf{79.7}
&
+15.72
&
74.2

\\

VGG16
&
2
&
1
&
66.07
&
\textbf{72.98}
&
+6.91
&
31.7

\\

MobileNetV2
&
3
&
1
&
70.28
&
\textbf{74.59}
&
+4.3
&
30.5

\\

\midrule

\textbf{Average} & 2.6 & 1.4 & 47.49 & \textbf{57.30} & +9.81 & 36.86 \\

\bottomrule

\end{tabular}
\end{adjustbox}

\end{table*}

\subsubsection{Impact on Reliability-Aware Partition Optimization}

Table~\ref{tab:critical_layers} summarizes how the identified
vulnerability-critical layers influence the final deployment strategy.
Across the evaluated networks, RAPID-DNN reassigns only a subset of
the identified critical layers relative to CNNParted, while the
remaining layer assignments are preserved. This demonstrates that the
optimizer selectively relocates only those layers for which the
expected reliability improvement justifies the associated latency and
energy overhead.
RAPID-DNN achieves an average reduction of 36.86\% in
fault-induced accuracy degradation while modifying only a small number
of layer assignments. These results indicate that deployment
reliability is primarily governed by a limited set of
vulnerability-critical layers rather than the complete network,
allowing robustness to be improved without extensive repartitioning.

\begin{figure*}[t]

\centering

\includegraphics[width=\textwidth]{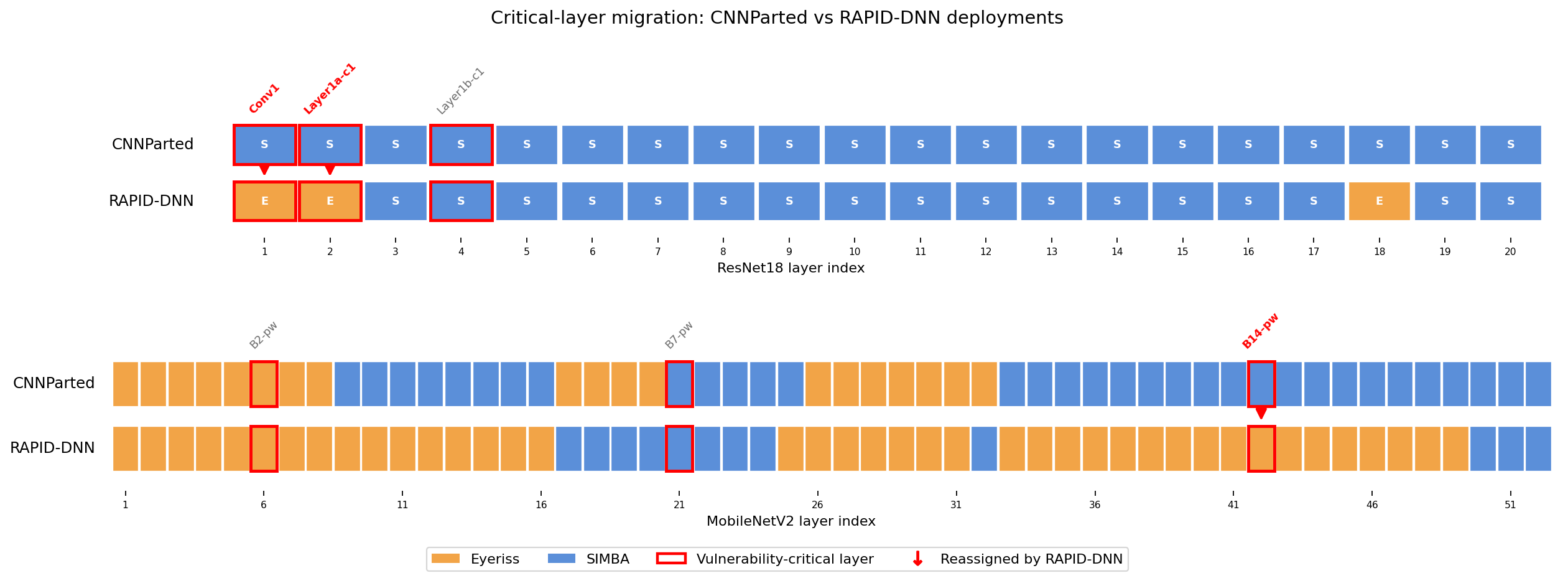}

\caption{Representative partition migration produced by RAPID-DNN for ResNet18 and MobileNetV2. Red boxes indicate vulnerability-critical layers identified during layer-wise reliability characterization. Downward arrows denote layers reassigned to different accelerators relative to CNNParted. Not all critical layers are migrated; reassignment occurs only when the expected reliability benefit outweighs the associated latency and energy cost, illustrating the multi-objective nature of the proposed optimization.}

\label{fig:layer_migration}

\end{figure*}

Figure~\ref{fig:layer_migration} illustrates representative partition
migrations for ResNet18 and MobileNetV2. The highlighted layers
correspond to the vulnerability-critical layers identified during the
characterization stage. Compared with CNNParted, RAPID-DNN places selected
vulnerability-critical layers on different accelerators when doing so
results in a better latency–energy–reliability trade-off during
multi-objective optimization. 
Across the evaluated networks, RAPID-DNN identifies two to
three vulnerability-critical layers, of which only one or two
are reassigned relative to CNNParted. 
% Even these modest
% partition modifications improve Top-1 accuracy by
% 3.61-18.50 percentage points while reducing accuracy
% degradation by 5.7-74.2\%.

The proposed layer-wise reliability characterization
successfully identifies the layers that dominate fault-induced
accuracy degradation for each network. Incorporating this information
into the optimization process enables RAPID-DNN to improve deployment
robustness by guiding the partition optimization toward
reliability-aware layer assignments rather than assuming uniform layer
vulnerability. 
These results demonstrate the effectiveness of the proposed layer-wise
reliability characterization in identifying vulnerability-critical DNN
layers.

%-----------------------------------------------------------
\subsection{Robustness Comparison Under Diverse Fault Conditions}
\label{sec:robustness}
%===========================================================

This subsection evaluates the robustness of the partitioning strategies
generated by RAPID-DNN under diverse runtime hardware fault conditions. Comparisons with the fault-agnostic CNNParted baseline are
performed under the same experimental configuration. 
The following
experiments examine the impact of increasing fault injection
probability and different bit-level fault models on inference
accuracy.

\subsubsection{Fault Injection Probability Analysis}
\label{subsec:fault_probability}

\begin{figure*}[t]
\centering
\includegraphics[width=\textwidth]{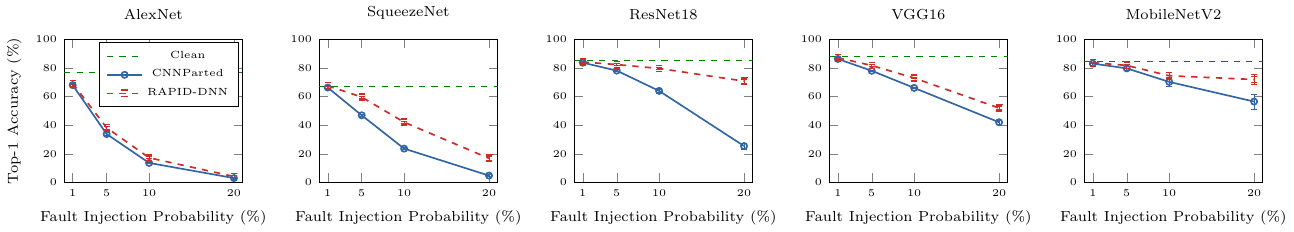}
\caption{Top-1 inference accuracy of CNNParted and RAPID-DNN under
increasing fault injection probabilities. Error bars denote 95\%
confidence intervals over $N=50$ independent stochastic
fault-injection trials.}
\label{fig:fault_rate}
\end{figure*}

Figure~\ref{fig:fault_rate} compares the Top-1 inference accuracy of
CNNParted and RAPID-DNN over fault injection probabilities ranging
from 1\% to 20\%, thereby evaluating the robustness of both
partitioning strategies under progressively increasing runtime fault
severity.
As expected, inference accuracy decreases
monotonically for both methods with increasing fault rates. However,
RAPID-DNN consistently maintains higher accuracy across all evaluated
networks and fault conditions.
The improvement becomes progressively more pronounced as the fault
injection probability increases. At 1\% fault rate, the Top-1 accuracy
improvement over CNNParted is limited to 0.48-1.26 percentage points
across the evaluated networks. At 5\%, the improvement increases to
2.08-12.74 percentage points, with SqueezeNet exhibiting the largest
gain. At 10\%, RAPID-DNN achieves improvements ranging from
3.61 percentage points for AlexNet to 18.50 percentage points for
SqueezeNet, while ResNet18 improves by 15.72 percentage points.
Under the most severe fault condition (20\%), the advantage becomes
even larger, reaching 45.46 percentage points for ResNet18,
20.48 percentage points for SqueezeNet, 15.94 percentage points for
MobileNetV2, 9.94 percentage points for VGG16, and 1.12 percentage
points for AlexNet.

Among the evaluated networks, ResNet18 exhibits the highest fault
tolerance under reliability-aware partitioning. Even at a 20\% fault
injection probability, RAPID-DNN maintains a Top-1 accuracy of
70.84\%, compared with only 25.38\% for CNNParted, recovering more
than two-thirds of the lost accuracy. Similar trends are observed for
SqueezeNet and MobileNetV2, whose accuracy remains substantially
higher than the baseline at elevated fault rates. These results
demonstrate that the benefit of reliability-aware partitioning
increases as runtime fault severity grows, enabling RAPID-DNN to
maintain significantly higher inference accuracy than the fault-
agnostic baseline under increasingly adverse operating conditions.

\begin{table*}[htbp]
\caption{Top-1 inference accuracy (\%) of CNNParted and RAPID-DNN under
different bit-level fault models. Gain denotes the absolute improvement
(percentage points) achieved by RAPID-DNN over CNNParted. The best
accuracy for each fault model is shown in bold. All results use a
10\% fault injection probability. Tables~\ref{tab:tradeoff}
and~\ref{tab:ablation} use the same setting.}
\label{tab:bitwise_results}

\centering
\footnotesize
\renewcommand{\arraystretch}{1.15}
\setlength{\tabcolsep}{3.5pt}

\begin{adjustbox}{max width=\textwidth}
\begin{tabular}{lcccccccccccc}

\toprule

\multirow{2}{*}{\textbf{Network}}
&
\multicolumn{3}{c}{\textbf{LSB}}
&
\multicolumn{3}{c}{\textbf{Higher-Order}}
&
\multicolumn{3}{c}{\textbf{Sign Bit}}
&
\multicolumn{3}{c}{\textbf{Random}}

\\

\cmidrule(lr){2-4}
\cmidrule(lr){5-7}
\cmidrule(lr){8-10}
\cmidrule(lr){11-13}

&
CNN
&
RAPID
&
Gain
&
CNN
&
RAPID
&
Gain
&
CNN
&
RAPID
&
Gain
&
CNN
&
RAPID
&
Gain

\\

\midrule

AlexNet
&
77.15
&
77.15
&
0.00
&
40.57
&
\textbf{52.00}
&
11.43
&
5.92
&
\textbf{7.79}
&
1.87
&
13.54
&
\textbf{17.15}
&
3.61

\\

SqueezeNet
&
67.35
&
\textbf{67.36}
&
0.01
&
18.14
&
\textbf{42.20}
&
24.06
&
35.53
&
\textbf{51.05}
&
15.52
&
23.57
&
\textbf{42.06}
&
18.49

\\

ResNet18
&
85.14
&
\textbf{85.16}
&
0.02
&
67.77
&
\textbf{81.95}
&
14.18
&
57.89
&
\textbf{77.34}
&
19.45
&
63.98
&
\textbf{79.70}
&
15.72

\\

VGG16
&
87.89
&
87.89
&
0.00
&
74.36
&
\textbf{79.81}
&
5.45
&
57.83
&
\textbf{66.22}
&
8.39
&
66.07
&
\textbf{72.98}
&
6.91

\\

MobileNetV2
&
84.34
&
\textbf{84.36}
&
0.02
&
72.75
&
\textbf{76.23}
&
3.48
&
74.69
&
\textbf{78.82}
&
4.13
&
70.28
&
\textbf{74.59}
&
4.31

\\

\midrule

\textbf{Average}
&
80.38
&
\textbf{80.38}
&
0.01
&
54.72
&
\textbf{66.44}
&
\textbf{11.72}
&
46.37
&
\textbf{56.24}
&
\textbf{9.87}
&
47.49
&
\textbf{57.30}
&
\textbf{9.81}

\\

\bottomrule

\end{tabular}
\end{adjustbox}

\end{table*}

%===========================================================
\subsubsection{Effect of Bit-Level Fault Models}
\label{subsec:bit_significance}
%===========================================================

The previous experiment investigated the influence of fault injection
probability on deployment robustness. This subsection evaluates the
robustness of the resulting partitions under different categories of
bit-level hardware faults. Candidate partitions are evaluated using the
fault-injection methodology described in
Section~\ref{sec:method}, in which the identified
vulnerability-critical layers are explicitly included during reliability
evaluation while additional faults are sampled according to the runtime
fault model. Four representative bit-level fault models are considered:
least significant bit (LSB), higher-order magnitude bit, sign-bit, and
random bit corruption.

Table~\ref{tab:bitwise_results} compares the inference accuracy of
CNNParted and RAPID-DNN under four representative bit-level fault
models. LSB faults have minimal impact on inference, with both methods
achieving an average Top-1 accuracy of 80.38\%, indicating that
perturbations confined to the least significant bit rarely alter the
network output. Consequently, the average improvement obtained by
RAPID-DNN is only 0.01 percentage points.
In contrast, substantially larger improvements are observed for
higher-order, sign-bit, and random faults. Under higher-order bit
corruptions, RAPID-DNN increases the average Top-1 accuracy from
54.72\% to 66.44\%, corresponding to an improvement of
11.72 percentage points. Similarly, average accuracy improves from
46.37\% to 56.24\% under sign-bit faults and from 47.49\% to
57.30\% under random bit faults, yielding gains of 9.87 and
9.81 percentage points, respectively. The largest per-network
improvements are observed for ResNet18 and SqueezeNet under the
higher-order and sign-bit fault models.

\begin{figure*}[t]
\centering
\includegraphics[width=\textwidth]{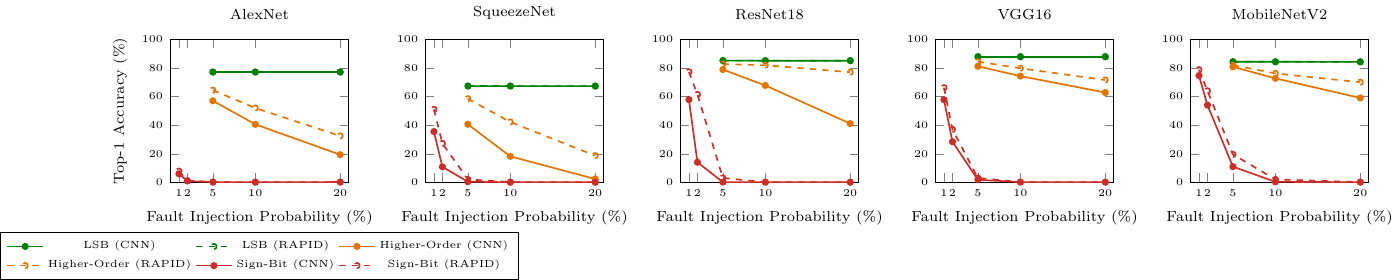}
\caption{Top-1 inference accuracy under different bit-level fault
models. Higher-order and sign-bit faults produce substantially larger
accuracy degradation than LSB faults, while RAPID-DNN consistently
maintains higher inference accuracy than CNNParted across all evaluated
fault categories.}
\label{fig:bit_significance}
\end{figure*}

Figure~\ref{fig:bit_significance} visualizes the network-wise accuracy
trends across the evaluated fault models. While RAPID-DNN and
CNNParted exhibit nearly identical performance under LSB faults, the
performance gap increases consistently for higher-order, sign-bit, and
random faults across all evaluated DNNs. The improvement is most
pronounced for ResNet18 and SqueezeNet, whereas AlexNet exhibits
comparatively smaller gains. This trend indicates that the advantage of
RAPID-DNN becomes increasingly significant as the numerical impact of
the injected faults increases.

RAPID-DNN consistently maintains or improves inference
accuracy across all evaluated bit-level fault models without degrading
performance under low-impact fault conditions. The largest gains are
observed for higher-order, sign-bit, and random faults, indicating
that the proposed layer-wise reliability characterization enables the
optimizer to identify deployment strategies that effectively mitigate
high-severity runtime faults while preserving accuracy under benign
fault conditions. These results demonstrate the robustness and generality of RAPID-DNN
across diverse bit-level fault models.

% %===========================================================
%===========================================================

\begin{table*}[htbp]
\caption{Latency--energy--reliability tradeoff obtained using the
proposed sensitivity-guided partition optimization.
CNNParted optimizes latency and energy, whereas RAPID-DNN additionally
optimizes expected fault-induced accuracy degradation.
Accuracy corresponds to Top-1 inference under the mixed random
bit-flip fault model described in Section~\ref{sec:method}, evaluated
at a 10\% fault injection probability. Percentage overheads are
computed relative to CNNParted.}
\label{tab:tradeoff}
\centering
\footnotesize
\renewcommand{\arraystretch}{1.15}
\setlength{\tabcolsep}{5pt}

\begin{adjustbox}{max width=\textwidth}
\begin{tabular}{lccccccccc}
\toprule
\multirow{2}{*}{\textbf{Network}}
&
\multicolumn{2}{c}{\textbf{Latency (ms)}}
&
\textbf{Latency}
&
\multicolumn{2}{c}{\textbf{Energy (mJ)}}
&
\textbf{Energy}
&
\multicolumn{2}{c}{\textbf{Top-1 Accuracy (\%)}}
\\

\cmidrule(lr){2-3}
\cmidrule(lr){5-6}
\cmidrule(lr){8-9}

&
CNNParted
&
RAPID-DNN
&
OH (\%)
&
CNNParted
&
RAPID-DNN
&
OH (\%)
&
CNNParted
&
RAPID-DNN
\\

\midrule

AlexNet
& 15.14 & 16.6 & 9.65
& 8.291 & 9.08 & 9.52
& 13.54 & \textbf{17.15}
\\

SqueezeNet
& 6.96 & 7.99 & 14.8
& 1.581 & 1.728 & 9.3
& 23.57 & \textbf{42.06}
\\

ResNet18
& 36.19 & 36.45 & 0.72
& 6.568 & 6.666 & 1.49
& 63.98 & \textbf{79.70}
\\

VGG16
& 302.72 & 309.6 & 2.27
& 61.91 & 63.609 & 2.74
& 66.07 & \textbf{72.98}
\\

MobileNetV2
& 43.12 & 43.09 & -0.06
& 4.529 & 4.563 & 0.75
& 70.28 & \textbf{74.59}
\\

\midrule

\textbf{Average}
& 80.82
& 82.75
& \textbf{2.38}
& 16.576
& 17.129
& \textbf{3.34}
& 47.49
& \textbf{57.30}
\\

\bottomrule
\end{tabular}
\end{adjustbox}

\smallskip
\begin{minipage}{\linewidth}\footnotesize
Average latency and energy overheads are computed from the aggregate
latency and energy across all evaluated networks rather than as the
arithmetic mean of the per-network overhead percentages.\\
\end{minipage}

\end{table*}

%===========================================================

\subsection{Latency--Energy--Reliability Tradeoff}
\label{sec:tradeoff}

Table~\ref{tab:tradeoff} summarizes the latency--energy--reliability
trade-off achieved by RAPID-DNN under the representative mixed random
bit-flip fault model. Compared with CNNParted, RAPID-DNN improves the
average Top-1 inference accuracy from 47.49\% to 57.30\%, corresponding
to an absolute gain of 9.81 percentage points, while incurring only
2.38\% average latency overhead and 3.34\% average energy overhead.
The overhead varies across DNN architectures depending on the selected
partition configuration. The largest latency overhead is observed for
SqueezeNet (14.80\%), whereas MobileNetV2 exhibits a slight latency
reduction (-0.06\%), indicating that improved reliability does not
necessarily require longer execution time. The energy overhead
ranges from 0.75\% (MobileNetV2) to 9.52\% (AlexNet), reflecting the
reassignment of vulnerability-critical layers to more fault-resilient
accelerator configurations.

%===========================================================

\begin{figure}[t]
\centering
\includegraphics[width=\columnwidth,height=4cm]{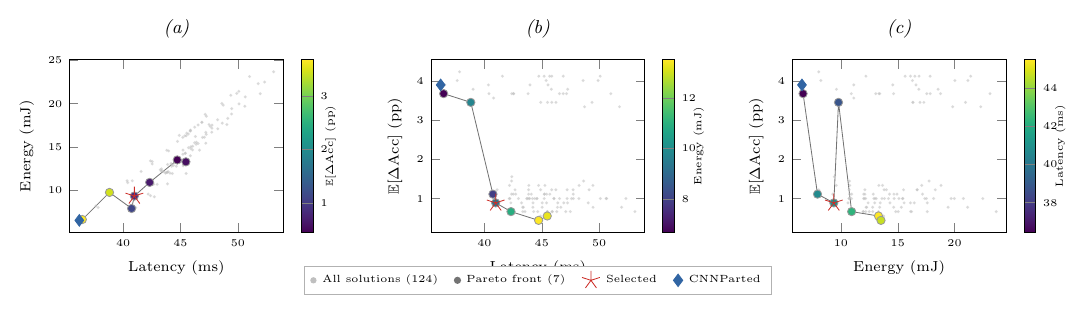}
\caption{Three-objective optimization results for ResNet18. Gray points represent all evaluated partition configurations, while the colored Pareto-optimal points and the connecting black line denote the Pareto front identified by NSGA-II. The red star indicates the selected deployment used in the subsequent evaluation, and the blue diamond denotes the CNNParted baseline. The three subplots show pairwise projections of latency, energy, and expected fault-induced accuracy degradation.}
\label{fig:pareto}
\end{figure}

Figure~\ref{fig:pareto} illustrates the multi-objective optimization
results for ResNet18. The gray markers represent the 124 feasible
partition configurations explored by NSGA-II, while the black curve
highlights the Pareto front comprising seven non-dominated solutions.
The three projections visualize the trade-offs among inference latency,
energy consumption, and expected fault-induced accuracy degradation. The
blue diamond denotes the CNNParted baseline, whereas the red star
indicates the Pareto-optimal deployment selected by RAPID-DNN. This
solution minimizes the expected accuracy degradation while satisfying
the imposed latency ($\leq$15\%) and energy ($\leq$50\%) overhead
constraints relative to CNNParted, and is therefore used throughout the
remaining experiments.

RAPID-DNN improves runtime robustness with only modest latency and
energy overheads, highlighting the effectiveness of reliability-aware
partitioning for AI deployments on heterogeneous edge accelerators.

\subsection{Ablation Study}
\label{sec:ablation}

Table~\ref{tab:ablation} compares three optimization strategies:
(i) CNNParted, which optimizes latency and energy without considering
runtime reliability; (ii) reliability-aware optimization with a uniform
layer sensitivity assumption (Uniform $S_l$); and (iii) the proposed
RAPID-DNN framework using measured layer-wise sensitivity scores
(Layer-wise $S_l$). All other experimental settings, including the DNN
models, hardware profiles, NSGA-II parameters, deployment constraints,
and fault-injection protocol, are kept identical. The Uniform $S_l$
configuration assigns the same normalized sensitivity score to every
layer ($S_l=1,\ \forall l$), whereas RAPID-DNN uses the measured
layer-wise sensitivity profile to guide partition optimization.

\begin{table*}[htbp]
\caption{Ablation study of RAPID-DNN under the random bit-flip fault model
at a 10\% fault injection probability. CNNParted performs conventional
latency--energy optimization without reliability awareness. Uniform~$S_l$
incorporates reliability as an optimization objective while assuming
identical layer sensitivity, whereas Layer-wise~$S_l$ represents the
complete RAPID-DNN framework using the proposed layer-wise reliability
characterization. Reported values are Top-1 inference accuracy (\%).}
\label{tab:ablation}
\centering
\footnotesize
\begin{adjustbox}{max width=\textwidth}
\begin{tabular}{lccc}
\toprule
\textbf{Network} & \textbf{CNNParted} & \textbf{Uniform $S_l$} & \textbf{Layer-wise $S_l$} \\
\midrule
AlexNet     & 13.54 & 17.15            & \textbf{17.15} \\
SqueezeNet  & 23.57 & 36.02 & \textbf{42.06} \\
ResNet18    & 63.98 & 78.21            & \textbf{79.70} \\
MobileNetV2 & 70.28 & 70.99 & \textbf{74.59} \\
VGG16       & 66.07 & 72.79            & \textbf{72.98} \\
\midrule
\textbf{Average} & 47.49 & 55.03 & \textbf{57.30} \\
\bottomrule
\end{tabular}
\end{adjustbox}
\end{table*}

Table~\ref{tab:ablation} summarizes the inference accuracy under the
random bit-flip fault model. Introducing a reliability objective with a
uniform layer sensitivity assumption (Uniform~$S_l$) improves the
average Top-1 accuracy from 47.49\% to 55.03\%, corresponding to an
absolute gain of 7.54 percentage points over the fault-unaware
CNNParted baseline. However, the Uniform~$S_l$ configuration exhibits
greater optimizer-seed variability for SqueezeNet and MobileNetV2,
indicating that assigning identical sensitivity to all layers provides
less stable optimization guidance.

In contrast, incorporating the proposed layer-wise reliability
information further increases the average Top-1 accuracy to
57.30\%, yielding an additional improvement of
2.27 percentage points over the Uniform~$S_l$ configuration
and an overall gain of 9.81 percentage points relative to
CNNParted. Compared with the Uniform~$S_l$ baseline, the
largest improvement is observed for SqueezeNet
(+6.04 percentage points), followed by MobileNetV2
(+3.60), ResNet18 (+1.49), and VGG16 (+0.19),
whereas AlexNet achieves identical accuracy under both
configurations.

These results show that incorporating runtime reliability into the
optimization process improves deployment robustness, while the proposed
layer-wise sensitivity characterization further enhances reliability by
guiding the optimizer toward more fault-tolerant partition
configurations.

\subsection{Reliability Estimation Convergence}
\label{sec:trial_convergence}

RAPID-DNN estimates the reliability objective using repeated
stochastic fault-injection trials. Consequently, the number of trials
directly influences both the statistical confidence of the estimated
expected accuracy degradation and the computational cost of the
optimization. This experiment evaluates the convergence of the
reliability estimate with respect to the number of fault-injection
trials.

\begin{figure}[t]
\centering
\includegraphics[width=\columnwidth]{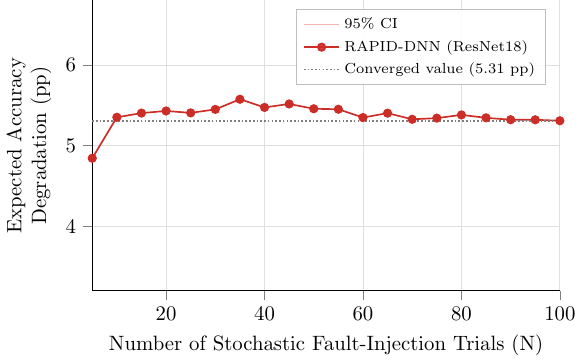}
\caption{Convergence of the estimated expected fault-induced accuracy
degradation with increasing numbers of stochastic fault-injection
trials. Shaded regions denote 95\% confidence intervals.}
\label{fig:trial_convergence}
\end{figure}

Figure~\ref{fig:trial_convergence} illustrates the convergence of the
estimated reliability metric for \textit{ResNet18} as a representative
example. As the number of stochastic fault-injection trials increases,
the estimated mean rapidly stabilizes, while the 95\% confidence
interval progressively narrows. Beyond approximately 40-50 trials,
both the estimated mean and confidence interval exhibit only minor
variation, indicating that additional trials provide diminishing
improvements in estimation accuracy. Similar convergence behavior was
observed for the other evaluated DNN models. Based on these
observations, all experiments in this work use $N=50$ fault-injection
trials, providing a practical balance between statistical stability and
computational cost.

\subsection{Discussion}

RAPID-DNN incorporates runtime reliability into the
partitioning process, enabling deployment strategies that improve
robustness against hardware faults while incurring only modest latency
and energy overheads. Unlike conventional performance-driven
partitioning, it exploits the non-uniform fault sensitivity of DNN
layers to guide partition selection. The ablation study shows that
introducing reliability as an optimization objective provides the
primary robustness improvement, while the proposed layer-wise
sensitivity characterization further refines partition decisions by
prioritizing vulnerability-critical layers. These results
highlight the importance of treating runtime reliability as a
co-optimization objective alongside latency and energy for DNN
deployment on heterogeneous edge accelerators.

Despite these encouraging results, the proposed framework has several
limitations. First, runtime reliability is evaluated using stochastic
bit-level fault injection rather than technology-specific device-level
fault models. Although this abstraction enables architecture-independent
evaluation, it does not capture the detailed electrical behavior of
fault mechanisms such as voltage glitches, electromagnetic
interference, and process variation. Extending the framework with
physics-based fault models is left for future work.

Second, RAPID-DNN focuses on reliability-aware deployment planning
rather than runtime fault detection, recovery, or hardware
implementation. Hardware latency and energy are estimated using
Timeloop and Accelergy, consistent with the evaluation methodology
adopted by CNNParted. The reported improvements reflect
the proposed optimization methodology under a common architectural
evaluation framework. However, analytical performance models cannot
capture all implementation-specific effects, including routing
congestion, timing closure, and memory-system behavior, that may
influence execution on physical FPGA platforms.

Finally, this work focuses on convolutional neural network inference on
heterogeneous edge accelerators. Future work will extend the proposed
methodology to transformer-based architectures, large language models,
and emerging heterogeneous AI accelerators, together with experimental
validation on physical FPGA-based platforms.

% ============================================================
\section{Conclusion}
\label{sec:conclusion}

RAPID-DNN introduces a reliability-aware DNN partitioning framework for
heterogeneous edge accelerators operating under runtime hardware
faults. By combining layer-wise reliability characterization with
multi-objective partition optimization, the framework jointly optimizes
inference latency, energy consumption, and runtime reliability.
Layer-wise sensitivity information is incorporated directly into the
optimization process, enabling partition decisions that improve
robustness while maintaining efficient execution.
Experimental evaluation on five CNN architectures using heterogeneous
Eyeriss and SIMBA accelerator profiles shows that RAPID-DNN
consistently improves inference robustness over the fault-agnostic
CNNParted baseline across different fault injection probabilities and
bit-level fault models, while incurring only modest latency and energy
overheads. The layer-wise sensitivity analysis, ablation study,
optimization convergence, and trade-off analysis further demonstrate the
effectiveness of incorporating runtime reliability into DNN partition
optimization.
Future work will focus on validation using physical heterogeneous
FPGA-based platforms, extending the framework to transformer-based
architectures and large language models, and investigating adaptive
runtime repartitioning under dynamically varying hardware fault
conditions.

\section*{Acknowledgments}
This work has received funding from the European Union's Horizon Europe
research and innovation programme under Grant Agreement No.~101225708, as
well as support from the Swiss State Secretariat for Education, Research and
Innovation (SERI).


\end{document}